\documentclass[aps,showpacs,prl,superscriptaddress,floatfix,twocolumn,citeautoscript,twocolumn]{revtex4-1}
\UseRawInputEncoding
\usepackage{graphicx}
\usepackage{amsmath}
\usepackage{amssymb}
\usepackage{bm}
\usepackage{dcolumn}
\usepackage{hyperref}
\usepackage[usenames]{color}
\usepackage{booktabs}
\usepackage{multirow}
\usepackage{siunitx}
\usepackage{wrapfig}
\usepackage{lipsum}
\usepackage{setspace}
\usepackage{tabularx}
\usepackage{booktabs}
\usepackage{framed}
\usepackage{xcolor}
\usepackage[normalem]{ulem}
\hypersetup{pdfborder=0 0 0,colorlinks=true,citecolor=blue,linkcolor=blue}
\input epsf

\newcommand\Tstrut{\rule{0pt}{2.6ex}} 
\usepackage{caption}
\usepackage{subcaption}
\newcommand{\editor}[2]{%
  \expandafter\newcommand\csname #1note\endcsname[1]{%
    \textcolor{#2}{(\textbf{#1:} ##1)}}%
  \expandafter\newcommand\csname #1\endcsname[1]{%
    \textcolor{#2}{##1}}%
  \expandafter\newcommand\csname #1cancel\endcsname[1]{%
    \textcolor{#2}{\sout{##1}}}%
  \expandafter\newcommand\csname #1change\endcsname[2]{%
    \textcolor{#2}{\sout{##1} ##2}}%
  \newenvironment{#1text}{\color{#2}}{\color{black}}
}

\usepackage{xr}
\makeatletter
\newcommand*{\addFileDependency}[1]{% argument=file name and extension
  \typeout{(#1)}
  \@addtofilelist{#1}
  \IfFileExists{#1}{}{\typeout{No file #1.}}
}
\makeatother

\usepackage{cleveref}

\crefrangeformat{equation}{Equations~#3#1#4--#5#2#6}
\crefmultiformat{equation}{Equations~#2#1#3}{ and~#2#1#3}{, #2#1#3}{ and~#2#1#3}
\crefformat{equation}{Equation~#2#1#3}
\Crefrangeformat{figure}{Figures~#3#1#4--#5#2#6}
\Crefmultiformat{figure}{Figures~#2#1#3}{ and~#2#1#3}{, #2#1#3}{ and~#2#1#3}
\Crefformat{figure}{Figure~#2#1#3}
\crefrangeformat{figure}{Figs.~#3#1#4--#5#2#6}
\crefmultiformat{figure}{Figs.~#2#1#3}{ and~#2#1#3}{, #2#1#3}{ and~#2#1#3}
\crefformat{figure}{Fig.~#2#1#3}
\crefrangeformat{table}{Tables~#3#1#4--#5#2#6}
\crefmultiformat{table}{Tables~#2#1#3}{ and~#2#1#3}{, #2#1#3}{ and~#2#1#3}
\crefformat{table}{Table~#2#1#3}
\crefname{section}{Section}{Sections}
\crefname{subsection}{Subsection}{Subsections}

\editor{IT}{rmagenta}
\editor{FH}{violet}
\editor{MG}{orange}

\editor{R}{red}

\begin{document}
\title{On-site and inter-site Hubbard corrections in magnetic monolayers: The case of FePS$_3$ and CrI$_3$}

\author{Fatemeh Haddadi} \email[]{ fatemeh.haddadi@epfl.ch}
\affiliation{Theory and Simulation of Materials (THEOS), \'Ecole Polytechnique F\'ed\'erale de Lausanne (EPFL), CH-1015 Lausanne, Switzerland}
\affiliation{National Centre for Computational Design and Discovery of Novel Materials (MARVEL), \'Ecole Polytechnique F\'ed\'erale de Lausanne (EPFL), CH-1015 Lausanne, Switzerland}

\author{Edward Linscott}
\affiliation{Theory and Simulation of Materials (THEOS), \'Ecole Polytechnique F\'ed\'erale de Lausanne (EPFL), CH-1015 Lausanne, Switzerland}

\author{Iurii Timrov}
\affiliation{Theory and Simulation of Materials (THEOS), \'Ecole Polytechnique F\'ed\'erale de Lausanne (EPFL), CH-1015 Lausanne, Switzerland}
\affiliation{National Centre for Computational Design and Discovery of Novel Materials (MARVEL), \'Ecole Polytechnique F\'ed\'erale de Lausanne (EPFL), CH-1015 Lausanne, Switzerland}

\author{Nicola Marzari}
\affiliation{Theory and Simulation of Materials (THEOS), \'Ecole Polytechnique F\'ed\'erale de Lausanne (EPFL), CH-1015 Lausanne, Switzerland}
\affiliation{National Centre for Computational Design and Discovery of Novel Materials (MARVEL), \'Ecole Polytechnique F\'ed\'erale de Lausanne (EPFL), CH-1015 Lausanne, Switzerland}

\author{Marco Gibertini}
\affiliation{Dipartimento di Scienze Fisiche, Informatiche e Matematiche, University of Modena and Reggio Emilia, I-41125 Modena, Italy}
\affiliation{Centro S3, CNR-Istituto Nanoscienze, I-41125 Modena, Italy}

\begin{abstract}
Hubbard-corrected density-functional theory has proven to be successful in addressing self-interaction errors in 3D magnetic materials. However, the effectiveness of this approach for 2D magnetic materials has not been extensively explored. Here, we use PBEsol+\emph{U} and its extensions PBEsol+\emph{U}+\emph{V} to investigate the electronic, structural, and vibrational properties of 2D antiferromagnetic FePS$_3$ and ferromagnetic CrI$_3$, and compare the monolayers with their bulk counterparts. Hubbard parameters (on-site \emph{U} and inter-site \emph{V}) are computed self-consistently using density-functional perturbation theory, thus avoiding any empirical assumptions. We show that for FePS$_3$ the Hubbard corrections are crucial in obtaining the experimentally observed insulating state with the correct crystal symmetry, providing also vibrational frequencies in good agreement with Raman experiments. For ferromagnetic CrI$_3$, we discuss how a straightforward application of Hubbard corrections worsens the results and introduces a spurious separation between spin-majority and minority conduction bands. Promoting the Hubbard \emph{U} to be a spin-resolved parameter --- that is, applying different (first-principles) values to the spin-up and spin-down manifolds --- recovers a more physical picture of the electronic bands and delivers the best comparison with experiments.
\end{abstract}
\date{\today}

\maketitle

%------------------------------------------------------------------------------------

\section{Introduction} 
Spintronic devices that exploit magnetic multi-layers are the backbone of modern technologies for magnetic sensing, data processing, and storage. The pursuit of novel magnetic materials with improved interfacial properties and decreasing thickness is still one of the main goals for spintronics studies, with van der Waals (vdW) materials holding great promise as they offer a versatile platform for exploring novel phenomena and provide high-quality interfaces at the atomic scale~\cite{sierra_van_2021,ahn_2d_2020}. However, magnetic applications for memories and processing were considered out of reach in vdW heterostructures, since magnetism had long been believed to hardly survive in two-dimensional (2D) systems because of the enhanced thermal fluctuations, as stated by the Mermin-Wagner theorem~\cite{mermin_absence_1966}. The recent discovery of 2D magnetic crystals~\cite{gong_discovery_2017, huang_layer-dependent_2017} brought experimental evidence that magnetic anisotropy can stabilize long-range magnetic order~\cite{onsager_crystal_1944}.
Such breakthrough opened the door to incorporating 2D magnetic materials in vdW heterostructures and spintronics devices~\cite{burch_magnetism_2018,gong_two-dimensional_2019,gibertini_magnetic_2019,mak_probing_2019,huang_emergent_2020,cortie_two-dimensional_2020}. For example, giant tunneling magnetoresistance has been observed in 2D magnets~\cite{song_giant_2018,klein_probing_2018,wang_very_2018,kim_one_2018}, which is promising for data-storage devices. The quest for high-density and low-energy-consumption devices (such as racetrack memories~\cite{parkin_memory_2015,tomasello_strategy_2015}) motivated the discovery of chiral spin textures, such as topologically protected skyrmions at room temperature in 2D magnets~\cite{park_ne-type_2021,wu_ne-type_2020,zhang_family_2021}. There is also an increasing interest in gate-tunable room-temperature magnetism~\cite{deng_gate-tunable_2018} and in controlling magnetism by electric fields~\cite{huang_electrical_2018,jiang_electric-field_2018} and currents~\cite{wang_current-driven_2019}. Furthermore, magnons in 2D magnets~\cite{cenker_direct_2021,xing_magnon_2019} could serve as a platform for wave-based computing technologies arising from magnon spintronics~\cite{chumak_magnon_2015}. All of these potential applications in  magnonics and spintronics may inherit the many advantages of 2D materials such as gate tunability, flexibility, low-cost, and large-scale growth~\cite{burch_magnetism_2018,gong_two-dimensional_2019,gibertini_magnetic_2019,mak_probing_2019,huang_emergent_2020,cortie_two-dimensional_2020}. In this context, understanding the physics governing in vdW magnets can be fruitful both for applications and further theoretical research. 

Density-functional theory (DFT) is a powerful tool to study the ground-state properties of materials~\cite{hohenberg_inhomogeneous_1964, kohn_self-consistent_1965, jones_density_2015} and there has been increasing interest to discover new 2D magnetic materials using DFT ~\cite{kabiraj_high-throughput_2020,torelli_high-throughput_2020,torelli_high_2019,mounet_two-dimensional_2018}. However, for magnetic materials with $d$- and/or $f$-shell electrons,  self-interaction errors (SIE) can be crucial~\cite{kulik_density_2006,cohen_insights_2008}. To address this problem, Hubbard corrections are often added to the DFT energy functional~\cite{anisimov_band_1991,anisimov_first-principles_1997,dudarev_electron-energy-loss_1998}, including an on-site Hubbard parameter \emph{U}~\cite{anisimov_band_1991} or even inter-site interactions \emph{V}~\cite{leiria_campo_jr_extended_2010}. However, determining the appropriate Hubbard parameters to be adopted in calculations is key. One approach is to use a semi-empirical on-site Hubbard \emph{U} chosen to reproduce some experimental data (e.g. band gaps, magnetic moments, oxidation enthalpies, etc.). However, this strategy is neither fully first-principles nor it can be applied to novel materials where experimental data are not available. A more systematic, parameter-free approach is to calculate Hubbard parameters self consistently, e.g.\ using linear-response theory~\cite{cococcioni_linear_2005}. In this scheme, the on-site Hubbard \emph{U} is chosen to restore a piece-wise linear behavior of the total energy with respect to the number of electrons in the Hubbard manifold~\cite{cococcioni_linear_2005}. This method has been recently streamlined through an efficient reformulation within density-functional perturbation theory (DFPT)~\cite{timrov_hubbard_2018, timrov_self-consistent_2021}, which is particularly useful for materials with localized partially-filled $d$- and/or $f$-shell electrons such as transition-metal and rare-earth compounds; however the current formulation is not applicable to electrons in closed $d$- and/or $f$-shells~\cite{yu_communication_2014}.

DFT+\emph{U}(+\emph{V}) calculations using Hubbard interactions parameters from DFPT have proven to be very effective in describing bulk 3D systems~\cite{floris_hubbard-corrected_2020,timrov_electronic_2020-1, Kirchner-hall_extensive_2021, mahajan_importance_2021, zhou_ab_2021, xiong_optimizing_2021, ricca_self-consistent_2020, mahajan_pivotal_2022-1, timrov_accurate_2022, timrov_unraveling_2023}. However, the importance of this method for 2D magnets is still an open question. Here we aim to explore this approach for 2D magnets,  investigating not only their electronic structure but also their vibrational properties, as these provide a reliable reference to compare first-principles results with Raman experiments. 
We focus on two particularly relevant and representative magnetic monolayers for which  Raman experimental data are available: antiferromagnetic FePS$_3$ ~\cite{lee_ising-type_2016,mccreary_distinct_2020,wang_raman_2020,du_weak_2016,ghosh_low-frequency_2021,vaclavkova_magnon_2021,mertens_ultrafast_2023} and ferromagnetic CrI$_3$~\cite{qiu_flexomagnetic_2023,huang_tuning_2020,kim_raman_2019,mccreary_distinct_2020,zhang_magnetic_2020}.

FePS$_3$ belongs to the family of transition-metal phosphorus trisulfides (MPS$_3$, M = Mn, Fe, Ni, ...). Other members of this family display clear signatures of magnetic ordering only down to bilayer systems, such as NiPS$_3$ and MnPS$_3$, with easy-plane and weak easy-axis magnetic anisotropy respectively ~\cite{kim_suppression_2019,kim_antiferromagnetic_2019}. However, magnetism in the monolayers is still controversial~\cite{long_persistence_2020}. On the contrary, FePS$_3$ has a strong out-of-plane anisotropy that suppresses thermal fluctuations and stabilizes magnetic ordering down to the monolayer limit~\cite{lee_ising-type_2016,lee_giant_2022}. Many first-principles studies have been performed on the electronic, magnetic, and vibrational properties of FePS$_3$~\cite{chittari_electronic_2016-1,lee_ising-type_2016,olsen_magnetic_2021-1,wang_raman_2016,amirabbasi_orbital_2023,deng_pressure-induced_2022,sheremetyeva_prediction_2023} with empirical values of Hubbard \emph{U} in the range between $2-7$~eV. Recent experiments show strong magnon-phonon coupling in FePS3 at high magnetic field, making this system promising for antiferromagnetic magnonics~\cite{liu_direct_2021, vaclavkova_magnon_2021, sun_magneto-raman_2022, cui_chirality_2023}.

In addition to FePS$_3$, we also study a ferromagnetic candidate: CrI$_3$, which is a member of the family of chromium trihalides (CrX$_3$, X~= Cl, Br, I, ...), which was the first 2D magnetic monolayer to be discovered experimentally~\cite{huang_layer-dependent_2017}. The other members of this family (such as CrCl$_3$ and CrBr$_3$) have shown in-plane and out-of-plane anisotropy axis with magnetism surviving down to the monolayer limit~\cite{chen_direct_2019,bedoya-pinto_intrinsic_2021}. The properties of CrI$_3$ have been extensively studied using DFT~\cite{soriano_magnetic_2020-1, kvashnin_dynamical_2022, ubrig_low-temperature_2019,pizzochero_inducing_2020,wu_physical_2019,ghosh_rashba-induced_2023}, with particular emphasis on the calculation of the exchange coupling constants~\cite{ke_electron_2021,kashin_orbitally-resolved_2020,pizzochero_magnetic_2020,besbes_microscopic_2019,lado_origin_2017,lu_curie_2019,wines_systematic_2023} as input for classical Monte Carlo simulations or analytical formulas~\cite{torelli_calculating_2018} aimed at extracting the Curie temperature. The vibrational properties of CrI$_3$ have also been studied via first principles~\cite{larson_raman_2018,lancon_magnetic_2016, djurdjic-mijin_lattice_2018,zhang_robust_2015,webster_distinct_2018}. In most cases, Hubbard corrections have either been neglected or introduced semi-empirically.

In this work, we study the structural, electronic, and vibrational properties of FePS$_3$ and CrI$_3$ fully from first principles using Hubbard functional and its extensions on top of PBEsol (i.e. PBEsol+$U$ and PBEsol+$U$+$V$). The on-site (\emph{U}) and inter-site (\emph{V}) interactions are computed self-consistently using DFPT as outlined in~\cite{timrov_hubbard_2018, timrov_self-consistent_2021} in a basis of L\"owdin-orthogonalized atomic orbitals~\cite{timrov_pulay_2020}, and later used to calculate phonon frequencies. Our findings show that  Hubbard corrections are essential to capture various properties of FePS$_3$ and CrI$_3$ in accordance with experiments. In the case of FePS$_3$, they play a crucial role in achieving the insulating ground state with the correct experimental symmetry, as well as in attaining good agreement with experimental phonon frequencies; the effects of Hubbard \emph{V} are relatively minor. In the case of CrI$_3$, while PBEsol already provides good structural and vibrational properties, the orbital character of the top of the valence bands is not correctly described. The Hubbard $U$ restores a correct picture for the valence bands  but at the same time gives rise to a spurious shift in the spin-minority conduction bands. While the introduction of Hubbard \emph{V} provides a slight improvement in the quantitative values of structural and vibrational properties, the incorrect positioning of the spin-minority conduction bands persists. This issue is effectively resolved through the implementation of spin-resolved $U$--different for spin-up and spin-down electrons, resulting in the overall best agreement with experimental data among the simulation strategies.

The paper is organized as follows. In \cref{sec2} we summarize the computational methods used in this work. In \cref{sec3} we show the results of ground-state properties for FePS$_3$ (\cref{subsec1}) and CrI$_3$ (\cref{subsec2}) from PBEsol, PBEsol+\emph{U}, and PBEsol+\emph{U}+\emph{V}. The results for the bulk structures are also provided for comparison. Finally, we provide our conclusions in \cref{sec4}. The phonon displacements and their corresponding frequencies, the description of the method to calculate the spin-resolved \emph{U}, and the analysis of the effects of Hund's exchange $J$ and of vdW interactions are provided in the Supplemental Material.

%-------------------------------------------------------------------------------------

\section{Computational details}
\label{sec2}
All calculations are performed using the \textsc{Quantum ESPRESSO} (QE) distribution~\cite{giannozzi_quantum_2009,giannozzi_advanced_2017,giannozzi_q_2020}. We use the exchange-correlation functional constructed using spin-polarized generalized-gradient approximation (GGA) with the PBEsol prescription~\cite{perdew_restoring_2008}. The values of 90 Ry (45 Ry) and 1080 Ry (360 Ry) have been set as the kinetic-energy cutoff for wavefunctions and spin-charge density, respectively, for FePS$_3$ (CrI$_3$) as suggested by the SSSP PBEsol library version 1.1.2~\cite{prandini_precision_2018,lejaeghere_reproducibility_2016,vanderbilt_soft_1990,dal_corso_pseudopotentials_2014,garrity_pseudopotentials_2014}. Unshifted $\mathbf{K}$ points meshes of size $6 \times 4 \times 1$ ($9 \times 9 \times 1$) for the monolayer and $4 \times 4 \times 6$ ($8 \times 8 \times 8$) for the bulk are used to sample the first Brillouin zone of of FePS$_3$ (CrI$_3$). In the monolayers, a vacuum of $17$~\AA\ is set in the direction perpendicular to the monolayer in order to ensure that the periodic images do not interact with each other. The results in the main text do not include vdW corrections as we are mainly interested in monolayers and their inclusion in bulk does not improve the agreement with experiments for the structural parameters (this is discussed further in Supplemental Material). Spin-orbit coupling is neglected in all calculations. The projected densities of states (PDOS) are plotted with a Gaussian broadening of 0.008~Ry.

Hubbard corrections are included in the calculations within the rotationally-invariant formalism of Dudarev \textit{et al.}~\cite{dudarev_electron-energy-loss_1998}; in Hubbard-corrected DFT, the total energy reads~\cite{leiria_campo_jr_extended_2010}:

\begin{equation}
    E_{\mathrm{DFT}+U+V} = E_{\mathrm{DFT}}+ E_{U+V} \,,
\end{equation}

where

\begin{eqnarray}
    E_{U+V}& = & \frac{1}{2} \sum_{I} \sum_{\sigma m_1m_2} U^I (\delta_{m_1m_2} - n^{I\sigma}_{m_1m_2}) n^{I\sigma}_{m_2m_1} \nonumber \\
    & & - \frac{1}{2} \sum_{I} \sum_{J(J\neq I)}^* \sum_{\sigma m_1m_2} V^{IJ} n^{IJ\sigma}_{m_1m_2} n^{IJ\sigma}_{m_2m_1} \,,
    \label{equation2}
\end{eqnarray}

where $I$ and $J$ are atomic site indices, $m_1$ and $m_2$ are the magnetic quantum numbers associated with a specific angular momentum, $U^I$ and $V^{IJ}$ are the on-site and inter-site Hubbard parameters respectively, and the star in the sum denotes that for each atom $I$, the index $J$ covers all its neighbors up to a given distance (or up to a given shell). The generalized atomic occupation matrices, $n^{IJ\sigma}_{m_1m_2}$, are computed by projecting the Kohn-Sham (KS) wavefunctions $\psi_{v\mathbf{k}\sigma}(\mathbf{r})$ on L\"owdin-orthogonalized atomic orbitals $\varphi_{m_1}^I(\mathbf{r})$ as: $n^{IJ\sigma}_{m_1m_2} = \sum_{v\mathbf{k}} f_{v\mathbf{k}\sigma} \langle \psi_{v\mathbf{k}\sigma} | \varphi_{m_2}^J \rangle \langle \varphi_{m_1}^I | \psi_{v\mathbf{k}\sigma} \rangle$, where $f_{v\mathbf{k}\sigma}$ are the occupations of KS states. Here, $v$ and $\sigma$ are the electronic band and spin indices, respectively, and $n^{I\sigma}_{m_1m_2} \equiv n^{II\sigma}_{m_1m_2}$. The magnetization of the $I$\textsuperscript{th} ion is calculated as $m^I = \sum_{m} \left( n^{I\uparrow}_{m m} - n^{I\downarrow}_{m m} \right)$. The equations above are for simplicity written in the framework of norm-conserving pseudopotentials and collinear polarization; the general formulation is discussed in Refs. \cite{timrov_hubbard_2018,timrov_self-consistent_2021}. We note that Eq.~\eqref{equation2} includes a double-counting term that corresponds to the ``fully localized limit (FLL)''~\cite{leiria_campo_jr_extended_2010}.

The Hubbard parameters, $U^I$ and $V^{IJ}$, are computed self-consistently using DFPT~\cite{timrov_hubbard_2018, timrov_self-consistent_2021} as implemented in the \textsc{HP} code~\cite{timrov_hp_2022}, which is part of \textsc{QE}. Importantly, computationally expensive summations over empty states in perturbation theory are avoided thanks to the use of projectors on empty states manifolds (see e.g. Refs.~\cite{baroni_phonons_2001, gorni_spin_2018}). The Hubbard parameters are defined as the diagonal and off-diagonal elements of the response matrices~\cite{cococcioni_linear_2005, leiria_campo_jr_extended_2010}: 

\begin{gather}
    U^{I} = (\chi^{-1}_{0} - \chi^{-1})_{II} , \\
    V^{IJ} = (\chi^{-1}_{0} - \chi^{-1})_{IJ} ,
\end{gather}

where $\chi$ and $\chi_{0}$ are the interacting and non-interacting response functions, respectively. The on-site Hubbard interactions ($U^{I}$) improve atomic-like localization on the Fe($3d$) (Cr($3d$)) states, while the inter-site Hubbard interactions ($V^{IJ}$) improve delocalizing covalent bonding between those states and the S($5p$) (I($3p$)) states. We use the self-consistent protocol for computing $U^{I}$ and $V^{IJ}$ as described in detail in Ref.~\cite{timrov_self-consistent_2021}.
The self-consistent procedure is initialized using the experimental crystal structure and zero Hubbard parameters. First, the Hubbard parameters are calculated using DFPT for the experimental structure; then, the Hubbard parameters are updated and the structure is optimized; in the next step, new Hubbard parameters are calculated for the relaxed structure and compared with the previous Hubbard parameters. This self-consistent process continues until the difference between the new and old Hubbard parameters becomes less than the convergence threshold (here, 0.01~eV).
For the calculation of Hubbard parameters, due to the relatively large unit cell for FePS$_3$, $\mathbf{q}$-points grids are set to $1 \times 1 \times 1$ for monolayer and bulk systems. We checked that by increasing the $\mathbf{q}$-points grids to $1 \times 1 \times 2$ in the bulk, the $U$ parameter changes by 0.06 eV, which is sufficiently small to have a negligible effect on the electronic structure. For CrI$_3$, $\mathbf{q}$-points grids of size $5 \times 5 \times 1$ and $3 \times 3 \times 3$ are used for monolayer and bulk, respectively. The self-consistent Hubbard parameters for CrI$_3$ and FePS$_3$ are given in \cref{t3}. We note that the $U$ value essentially does not change when we go from bulk to monolayer. This is an interesting observation, although within the current formulation based on linear response theory, it is not surprising as $U$ is calculated in order to impose piecewise linearity to the energy functional as electrons are added or subtracted to the very localized (``Hubbard'') manifold of $d$ electrons, and to remove self-interaction errors. As such, it is very weakly dependent on the $d$ electrons being in a monolayer or in the bulk of a vdW material, since the chemistry of the interaction between the transition-metal ion and the ligands is very much unaffected by the stacking. This is very different from what happens within an alternative first-principles method for estimating the Hubbard parameter: the so-called cRPA approach~\cite{aryasetiawan_frequency-dependent_2004}. Although the two approaches unfortunately share the same name of ``Hubbard $U$'', the $U_{\rm cRPA}$ that is computed within cRPA is driven by a completely different hypothesis, that is to calculate the average (on the $d$ orbitals of interest) partially-screened interaction towards a better description of the spectral properties of a (correlated) material. This is very much affected by the 2D or 3D environment, typically leading to an increase of $U_{\rm cRPA}$ from bulk to monolayer as a consequence of the suppression of screening from the adjacent layers~\cite{soriano_environmental_2021-1}.

\begin{table}[t]
    \renewcommand{\arraystretch}{1.5}
    \caption{The values of self-consistent Hubbard parameters \emph{U}~(eV) describing the strength of on-site interactions for Fe($3d$) and Cr($3d$) states, and \emph{V}~(eV) describing the strength of inter-site interactions for Fe($3d$)--S($3p$) and Cr($3d$)--I($5p$) couples, for bulk and monolayer FePS$_3$  and CrI$_3$.}
    \begin{tabularx}{\columnwidth}{X c *{2}{S[table-format=2.3]} *{2}{S[table-format=2.3]}}
       \hline
       \hline
       & & \multicolumn{2}{c}{FePS$_3$} & \multicolumn{2}{c}{CrI$_3$} \\
       &  & \multicolumn{1}{c}{bulk} & \multicolumn{1}{c}{monolayer} &  \multicolumn{1}{c}{bulk} & \multicolumn{1}{c}{monolayer} \\
        \hline
        PBEsol+\emph{U} & \emph{U} & 4.94 & 4.93 & 6.61 & 6.54\\
        \hline
        \multirow{2}{*}{PBEsol+\emph{U}+\emph{V}} & \emph{U} & 4.92 & 4.94 & 6.53 & 6.41\\
        \cline{2-6}
        & \emph{V} & 0.21 & 0.22 & 0.26 & 0.28\\
        \hline
        \hline
    \end{tabularx}
    \label{t3}
\end{table}

The phonon frequencies have been computed using the frozen-phonon method as implemented in the \textsc{Phonopy} package~\cite{togo_first_2015}. The phonon frequencies were calculated on a $2 \times 2 \times 2$ supercell for the bulk systems and on a  $2 \times 2 \times 1$ supercell for monolayer systems  (we tested for CrI$_3$ that no significant difference could be observed by increasing it to $3 \times 3 \times 3$. For FePS$_3$ due to the computational costs we were unable to test larger supercell.). For the monolayers, the corrections of translational, rotational invariance, and equilibrium conditions are applied on interatomic force constants (IFCs) to recover the correct quadratic behavior close to the Brillouin zone center of the ZA phonon band (i.e. out-of-plane transverse acoustic mode)~\cite{lin_general_2022}. Since we compare against available Raman experiments and in two dimensions the longitudinal optical and transverse optical (LO--TO) splitting breaks down at the $\Gamma$ point~\cite{sohier_breakdown_2017}, the non-analytical term of the dynamical matrix for monolayer systems is neglected. For bulk systems, the non-analytical corrections are included up to dipolar order through Born effective charges (BECs) and the dielectric tensor. To calculate BECs and the dielectric tensor, two approaches can be used: finite differences~\cite{umari_ab_2002,souza_maximally_2001} and DFPT~\cite{giannozzi_ab_1991, gonze_dynamical_1997, baroni_phonons_2001, tobik_electric_2004}. The current implementation of the DFPT method can only be applied for DFT+\emph{U} with non-orthogonalized atomic projections~\cite{floris_hubbard-corrected_2020}. In the Supplemental Material, we show for bulk FePS$_3$ that BECs using the first method with L\"owdin-orthogonalized Hubbard projections and the second method with non-orthogonalized atomic Hubbard projections give very similar results when the structure is the same. Therefore, we use DFPT with non-orthogonalized atomic Hubbard projectors to calculate BECs and the dielectric tensor due to the cheaper computational cost and convergence issues encountered when using the finite differences for bulk CrI$_3$.

The data used to produce the results of this paper are available in the \href{https://archive.materialscloud.org/record/2024.18}{\color{blue}Materials Cloud Archive}.

\section{Results and discussion}
\label{sec3}

%-------------------------------------------------------------------------------------

\subsection{FePS$_3$}
\label{subsec1}

\begin{figure}[h]
    \centering
    \captionsetup{justification=Justified,singlelinecheck=false}
    \includegraphics[width=1\columnwidth]{ 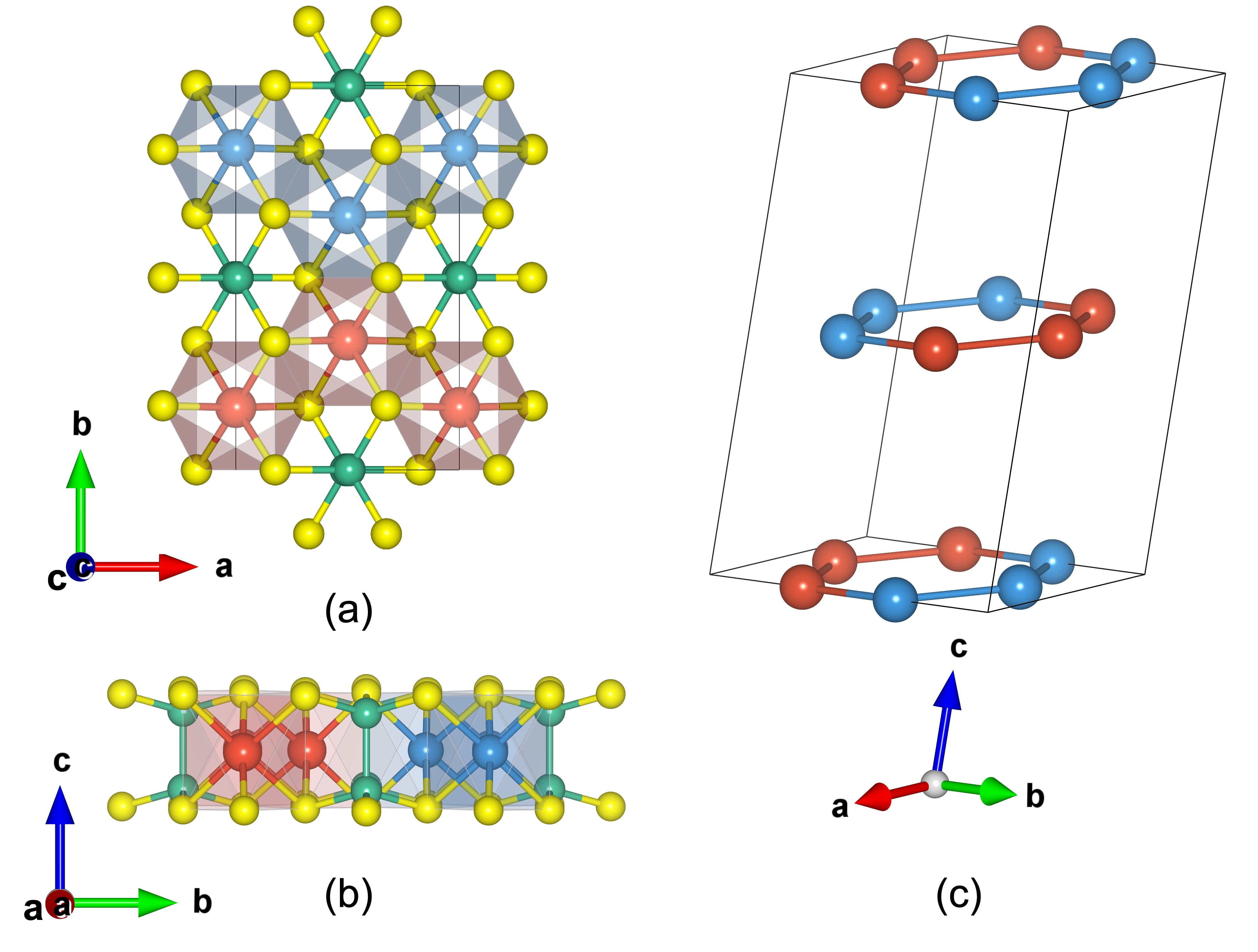}
    \caption{(a) Top view and (b) side view of monolayer FePS$_3$. The unit cell is shown in black lines. The red, blue, yellow, and green balls correspond to spin-up Fe, spin-down Fe, S, and P atoms, respectively. The magnetic configuration in the bulk system is shown in (c), where only magnetic (Fe) atoms are shown.
    The illustrations are obtained using  VESTA~\cite{momma_it_2011}.}
    \label{fig1}
\end{figure}

The first system that we study is the 2D antiferromagnetic Ising-type FePS$_3$. Figures~\ref{fig1}(a) and (b) show the top and side view of the FePS$_3$ monolayer. The Fe atoms form a planar honeycomb lattice and are enclosed in octahedra of six S atoms. These S atoms are also connected to two P atoms in the center of the Fe hexagons. The primitive unit cell of both monolayer and bulk systems contains 4 Fe, 4 P, and 12 S atoms. Bulk FePS$_3$ forms a monoclinic structure with the space group C2/m (No.\ 12) and point group C$_{2h}$\cite{ouvrard_structural_1985}.
In the ground state, FePS$_3$ is an antiferromagnet where, within each layer, Fe atoms are ferromagnetically ordered along  zigzag chains, but then each chain is antiferromagnetically aligned with respect to its neighbors~\cite{lancon_magnetic_2016} (\cref{fig1}(a)), possibly leading to a nematic state~\cite{ni_observation_2022}. In the bulk system, a further antiferromagnetic ordering involves zigzag chains in adjacent layers (\cref{fig1}(c)). Because of this magnetic configuration, the unit cell of the antiferromagnetic state in a monolayer is twice that of the ferromagnetic or nonmagnetic state. The doubled in-plane unit cell results in a halved first Brillouin zone~\cite{lee_ising-type_2016}.

\begin{figure*}[bt!]
        \captionsetup{singlelinecheck = false, justification=raggedright}
     \centering
     \begin{subfigure}[t]{.47\textwidth}
         \centering
          \caption{}
         \includegraphics[width=\textwidth]{ 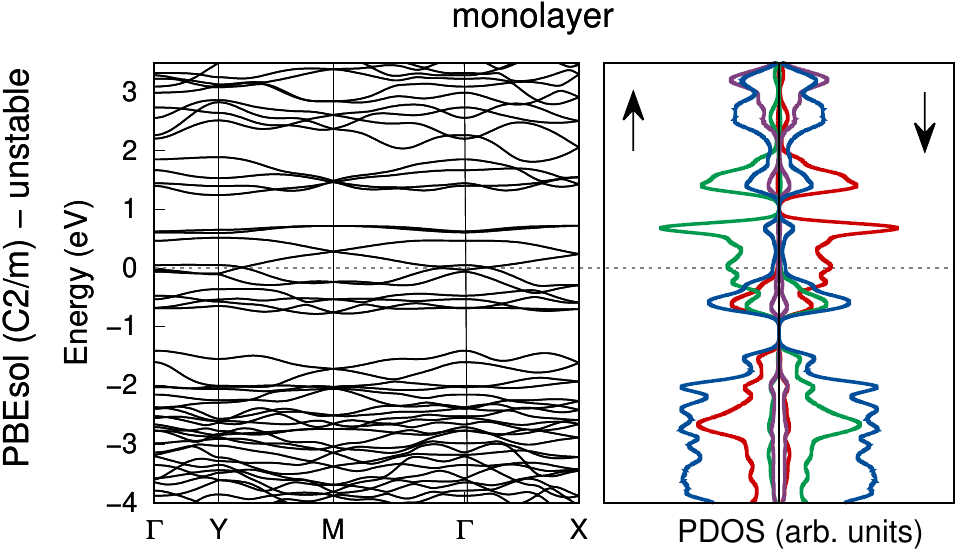}
         \vspace{-4ex}
         \label{fe-band-1}
     \end{subfigure}
     \hfill
     \begin{subfigure}[t]{.44\textwidth}
         \centering
         \caption{}
         \includegraphics[width=\textwidth]{ 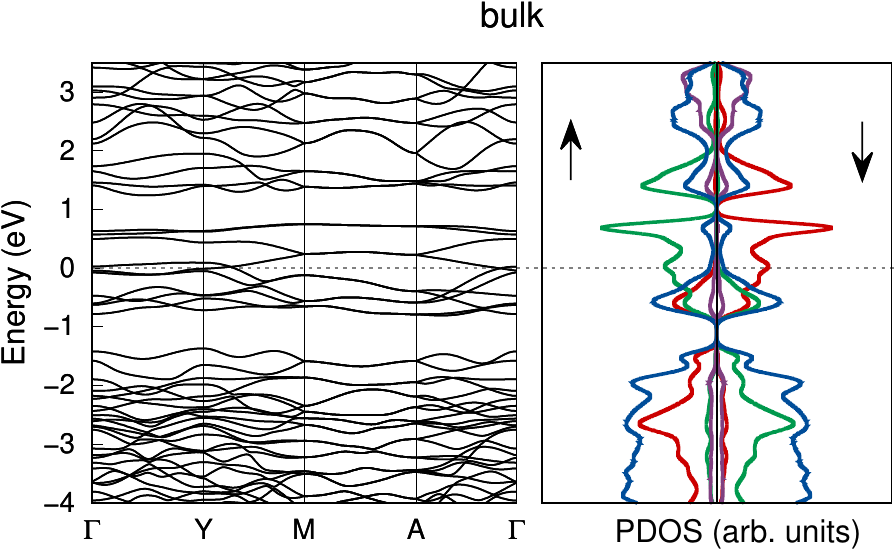}
         \vspace{-4ex}
         \label{fe-band-2}
     \end{subfigure}
          \newline
     \begin{subfigure}[t]{.47\textwidth}
         \centering
         \caption{}
         \includegraphics[width=\textwidth]{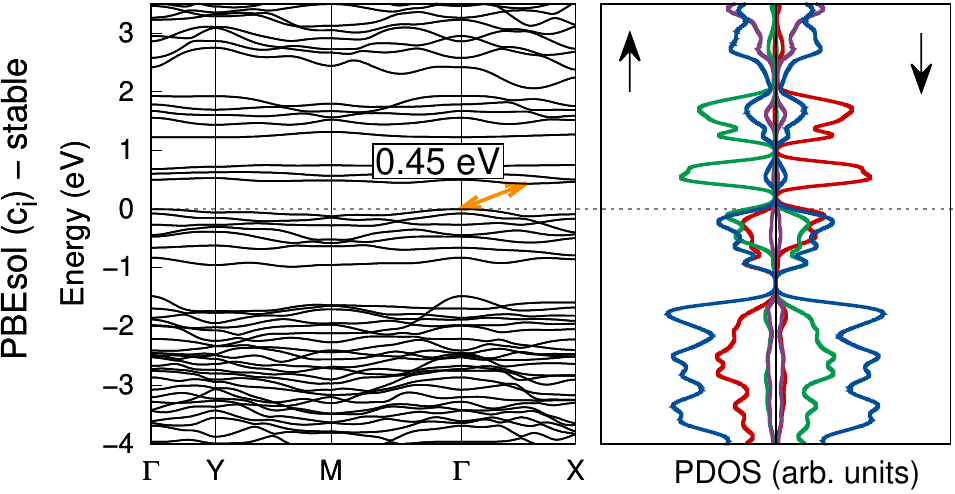}
         \vspace{-4ex}
         \label{fe-band-3}
     \end{subfigure}
     \hfill
          \begin{subfigure}[t]{.44\textwidth}
          \centering
          \caption{}
         \includegraphics[width=\textwidth]{ 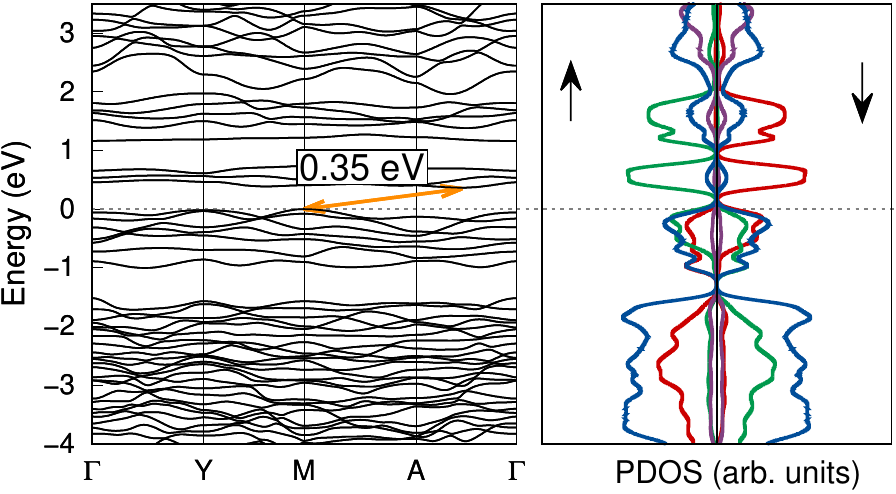}
         \vspace{-4ex}
         \label{fe-band-4}
     \end{subfigure}
     \newline
     \begin{subfigure}[t]{.47\textwidth}
         \centering
         \caption{}
         \includegraphics[width=\textwidth]{ 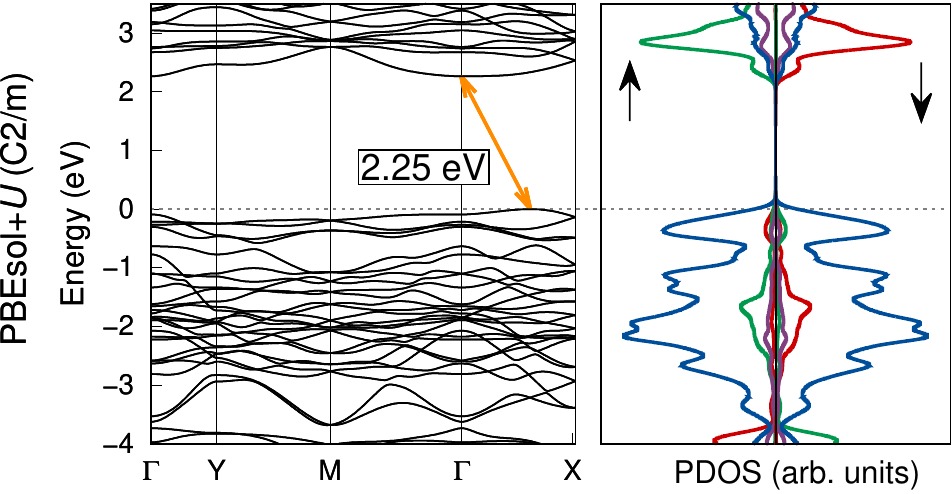}
         \vspace{-4ex}
         \label{fe-band-5}
     \end{subfigure}
          \hfill
     \begin{subfigure}[t]{.44\textwidth}
         \centering
         \caption{}
         \includegraphics[width=\textwidth]{ 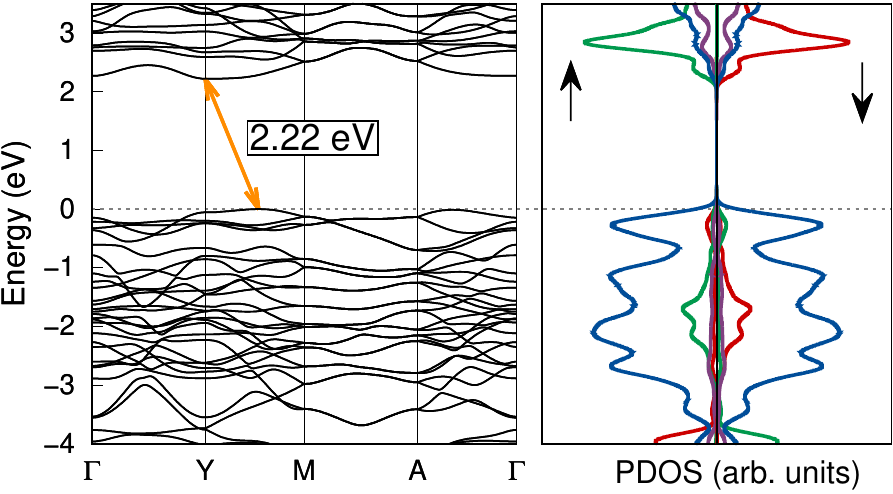}
         \vspace{-4ex}
         \label{fe-band-6}
     \end{subfigure}
          \newline
               \begin{subfigure}[t]{.47\textwidth}
         \centering
         \caption{}
         \includegraphics[width=\textwidth]{ 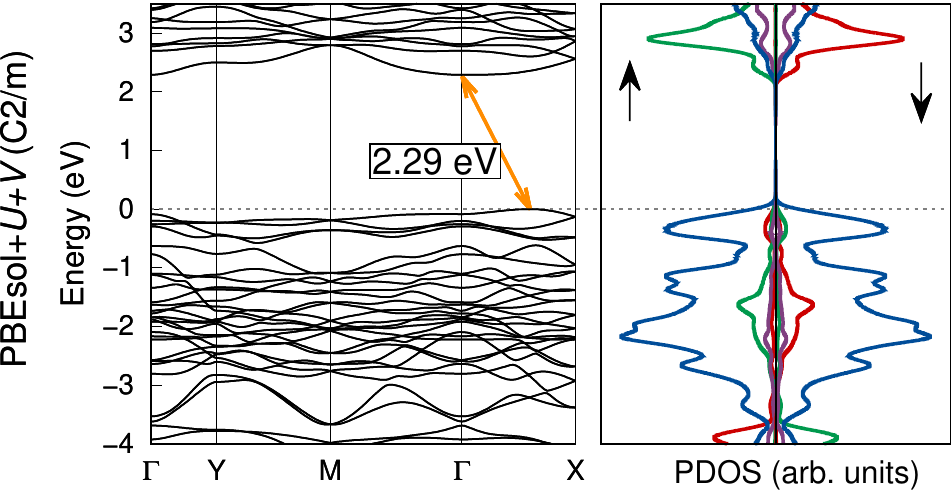}
         \vspace{-4ex}
         \label{fe-band-7}
     \end{subfigure}
          \hfill
     \begin{subfigure}[t]{.44\textwidth}
         \centering
         \caption{}
         \includegraphics[width=\textwidth]{ 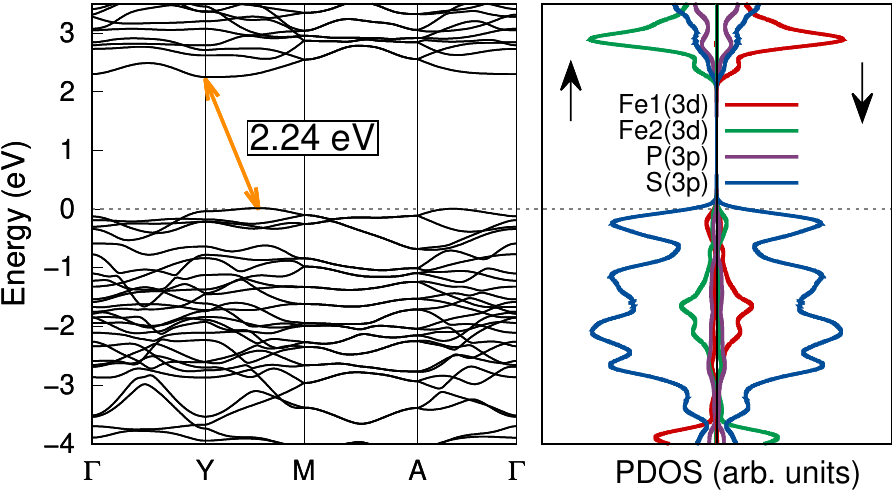}
         \label{fe-band-8}
     \end{subfigure}
    \captionsetup{justification=Justified}
     \caption{The electronic band structure and spin-resolved projected density of states (PDOS) of monolayer FePS$_3$ (first column) and bulk (second column) from (a) and (b) PBEsol (metallic, and unstable with C2/m space group symmetry), (c) and (d) PBEsol (stable with C$_i$ space group symmetry), (e) and (f) PBEsol+\emph{U} (C2/m), and (g) and (h) PBEsol+\emph{U}+\emph{V} (C2/m). Orange arrows indicate the indirect band gaps, and their values are also reported. Fe1 and Fe2 are iron atoms with up and down magnetization, respectively. The color guide is given in (h); the dashed line is the Fermi energy for (a) and (b), and the top of the valence bands in all other cases}
     \label{fe-band}
\end{figure*}

First, we show the electronic bands from PBEsol and PBEsol+\emph{U}(+\emph{V}) calculations in~\cref{fe-band}. During the self-consistent process of the calculation of Hubbard parameters, the crystal space group symmetry is constrained to the experimental symmetry, C2/m (No.\ 12). For the experimental symmetry, PBEsol predicts metallic behavior both for the monolayer and bulk (\cref{fe-band-1,fe-band-2}). However, these systems are unstable and acquire soft phonons, which can be removed by lifting the constraint to the experimental symmetry and optimizing the structure again. The optimized structure with C$_i$ (No.\ 2) symmetry has lower energy than the symmetric structure and does not show imaginary phonon frequencies. \Cref{fe-band-3,fe-band-4} show that the distorted systems are no longer metallic, although the band gap is still significantly smaller than in experiments (0.45 versus $2.18$~eV~\cite{cheng_high-yield_2018} for quantum sheets and 0.35 versus $0.5-1.6$~eV~\cite{haines_pressure-induced_2018,du_weak_2016,brec_physical_1979,foot_optical_1980} for the bulk). On the other hand, the PBEsol+\emph{U}(+\emph{V}) calculations (\cref{fe-band-5,fe-band-8}) predict an insulating and stable ground state with C2/m space group, thus preserving experimentally determined symmetry. For the bulk the band gap is overestimated; for the monolayer, the band gap is in very good agreement with experiments. The on-site Hubbard \emph{U} correction is thus crucial to recover the correct electronic structure and crystal symmetry of FePS$_3$. Including inter-site Hubbard interactions ($+V$) has a negligible effect, as shown in \cref{fe-band-7,fe-band-8}.

\begin{table*}[!hbt]
    \renewcommand{\arraystretch}{1.5}
    \caption{\label{t1} Lattice parameters $a$, $b$, and $c$ (in \AA) and absolute magnetization $m_{Fe}$ of Fe atoms (in $\mu_B$; calculated from $m^I = \sum_{m} \left( n^{I\uparrow}_{m m} - n^{I\downarrow}_{m m} \right)$ as described in section \ref{sec2}) computed using different approaches for monolayer and bulk FePS$_{3}$. For PBEsol, results are reported for the stable structure with reduced symmetry (\cref{fe-band-3,fe-band-4}). Experimental data for the bulk are from Refs.~\cite{wiedenmann_neutron_1981,lancon_magnetic_2016,kurosawa_neutron_1983}.}
    \centering
    \begin{tabularx}{\textwidth}{X X *{4}{D..{5.7}} D..{5.6}}
    \hline\hline
    & & \multicolumn{1}{c}{$a$} & \multicolumn{1}{c}{$b$} & \multicolumn{1}{c}{$b/\sqrt{3}a$} & \multicolumn{1}{c}{$c$} & \multicolumn{1}{c}{$|m_{Fe}|$} \\
    \hline
    \multirow{4}{*}{bulk}
    & Expt.                    & 5.947 & 10.300 & 1.000 & 6.722 & \multicolumn{1}{c}{4.9, $4.52\pm0.05$, $5.1\pm0.6$}  \\
    & PBEsol                   & 5.742 & 10.188 & 1.024 & 6.479 & 3.20 \\
    & PBEsol+\emph{U}          & 5.955 & 10.253 & 0.994 & 6.699 & 3.63 \\
    & PBEsol+\emph{U}+\emph{V} & 5.952 & 10.245 & 0.994 & 6.696 & 3.62 \\
    \hline
    \multirow{3}{*}{monolayer}
    & PBEsol                   & 5.745 & 10.184 & 1.023 & \multicolumn{1}{c}{---} & 3.25 \\
    & PBEsol+\emph{U}          & 5.954 & 10.251 & 0.994 & \multicolumn{1}{c}{---} & 3.63 \\
    & PBEsol+\emph{U}+\emph{V} & 5.950 & 10.243 & 0.994 & \multicolumn{1}{c}{---} & 3.62 \\
    \hline\hline
    \end{tabularx}
\end{table*}

We also note that the computed band gap barely changes when going from bulk to monolayer. This is the case also when considering vdW-compliant functionals on top of the same crystal structure (see Supplemental Material). This observation might appear in contrast with what typically happens in layered materials, where quantum confinement effects tend to significantly increase the band gap as thickness is reduced. We attribute the negligible band gap variation to a reduced interlayer hopping, resulting from 1) the fact that electronic states around the band gap (especially the conduction states) show a large contribution from $d$ orbitals, which are typically very localized, limiting tunneling between the layers and 2) the antiferromagnetic ordering between the layers. As a consequence, electronic states tend to be confined within a single layer even in the bulk and the change in band gap due to quantum confinement is reduced with respect to other layered materials. Still, in experiments~\cite{cheng_high-yield_2018} the band gap difference between bulk and so-called quantum sheets is about 0.6 eV (2.18 eV for the quantum sheets and 1.6 eV for the bulk). The discrepancy with our calculations might arise either from an uncertainty in the measurements, where the band gap is extracted from a linear extrapolation of the smeared spectrum of optical absorption at room temperature, or from aspects that are not captured in our calculations, such as a slight variation in the crystal structure or a more consistent description of screening and many-body effects. For instance, in Ref.~\cite{budniak_spectroscopy_2022}, the authors use first-principles calculations to obtain a band gap difference of 0.4 eV between monolayer and bulk, closer to the experiments, which is associated with a non-negligible variation in the in-plane lattice parameter.  Moreover, it is also important to remember that DFT is not expected to provide accurate predictions for band gaps (including their variation from bulk to monolayer). Hubbard corrections with the $U$ parameter computed from a piecewise-linearity condition (e.g. from linear response~\cite{cococcioni_linear_2005,timrov_self-consistent_2021, timrov_hubbard_2018}) often improve band gaps significantly, at least when band edges are dominated by states that are mostly those of the Hubbard manifold~\cite{Kirchner-hall_extensive_2021}. More systematic improvements may require advanced many-body approaches such as GW calculations~\cite{reining_gw_2018, golze_gw_2019-1, da_jornada_nonuniform_2017, rudenko_toward_2015, cheiwchanchamnangij_quasiparticle_2012} (due to their improved description of long-range Coulomb interactions and electronic screening), hybrid functionals~\cite{skone_self-consistent_2014, skone_nonempirical_2016, ohad_band_2022, yang_range-separated_2023, liu_assessing_2020, wing_band_2021}, or Koopmans functionals that correct the band gap by design~\cite{nguyen_koopmans-compliant_2018, linscott_koopmans_2023, colonna_koopmans_2022}. Therefore, more investigations of the band gap differences between bulk and monolayer FePS$_3$ are needed both on the experimental as well as on the theoretical side.

Regarding the instabilities seen in PBEsol, we note that a similar effect was also observed in Ref.~\cite{hashemi_vibrational_2017}, from PBE calculations. By choosing an empirical value of \emph{U} (3.5~eV) that yields the same energy difference between the FM and AFM configurations as obtained from hybrid HSE calculations, the authors showed that the Hubbard \emph{U} correction could almost completely remove the instability~\cite{hashemi_vibrational_2017}. Other studies~\cite{lee_ising-type_2016,kargar_phonon_2020} reported that the instability survives even in DFT+\emph{U} calculations (with  empirical values for \emph{U} of 4.2~eV in Ref.~\cite{lee_ising-type_2016} and 3.5~eV in Ref.~\cite{kargar_phonon_2020}, using projector-augmented wave (PAW) Hubbard projectors). Such instabilities seen in DFT+\emph{U} calculations could also be a result of the presence of multiple local minima in the total energy when considering Hubbard-corrected energy functionals, and the ensuing difficulty of finding the correct global energy minimum~\cite{meredig_method_2010}. One way to find the global minimum would be to start the calculations from different occupation matrices, as done in Ref.~\cite{amirabbasi_orbital_2023}.

\Cref{fe-band} also shows the spin-resolved projected density of states (PDOS) of FePS$_3$. 
We note that the Fe($3d$) states span a wide energy range when using plain PBEsol, but become more localized in Hubbard-corrected PBEsol. While in PBEsol there is a strong contribution at the top of the valence bands of the Fe($3d$) states with a strong hybridization with S($3p$) states (\cref{fe-band-3,fe-band-4}), the inclusion of Hubbard corrections pushes Fe($3d$) states down in energy and leaves the top of the valence bands dominated by S($3p$) states. The bottom of the conduction bands remains dominated by Fe($3d$) states for both approaches.

We now focus in detail on the structural properties and magnetization for FePS$_3$, as summarized in \cref{t1}. PBEsol+\emph{U}(+\emph{V}) shows better agreement with experiments than PBEsol, corroborating the results of Ref.~\cite{olsen_magnetic_2021-1} where the \emph{U} parameter was shown to be important to obtain the correct magnetic properties of $M$PS$_3$ ($M$ = Fe, Ni, Mn) materials. We can see in~\cref{t1} that, as expected, the magnetization increases after the inclusion of Hubbard corrections. Remarkably, the lattice parameters and magnetic moments do not vary much from monolayer to bulk, suggesting that these quantities are not dependent on the number of layers. From \cref{t1} it is also clear that a stronger monoclinic distortion ($b\neq\sqrt{3}a$) is predicted by the calculations than that observed in experiments. This distortion is accompanied by a different distance between Fe ions with the same spin orientation (up-up or down-down, given by $a/\sqrt{3}$) and with opposite spin orientation (up-down or down-up, given by $b/3$), with a value larger by about 0.02~\AA\ for parallel spins when Hubbard corrections are included, while the opposite distortion by 0.08~\AA\ is present at the PBEsol level. Last, \cref{t1} shows that the inclusion of Hubbard \emph{V} corrections change the lattice constants and magnetization negligibly, suggesting the inter-site interactions between Fe($3d$) and S($3p$) states do not have considerable effects on structural properties and magnetization.

\begin{figure*}[t]
      \captionsetup{singlelinecheck = false, justification=raggedright}
      \centering
      \begin{subfigure}[t]{.32\textwidth}
        \centering
        \caption{}
        \includegraphics[scale=0.75]{ 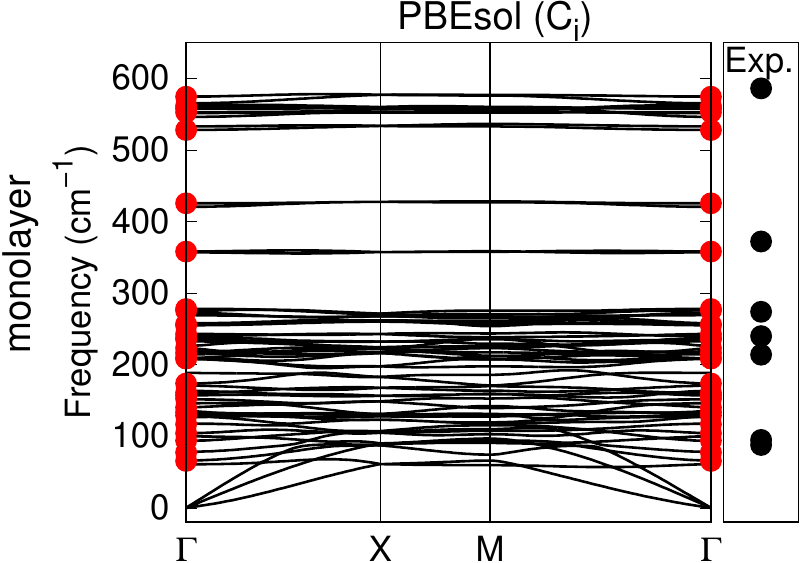}
        \vspace{-4ex}
        \label{fe-ph-1}
      \end{subfigure}
      \hfill
      \begin{subfigure}[t]{.3\textwidth}
        \centering
        \caption{}
        \includegraphics[scale=0.75]{ 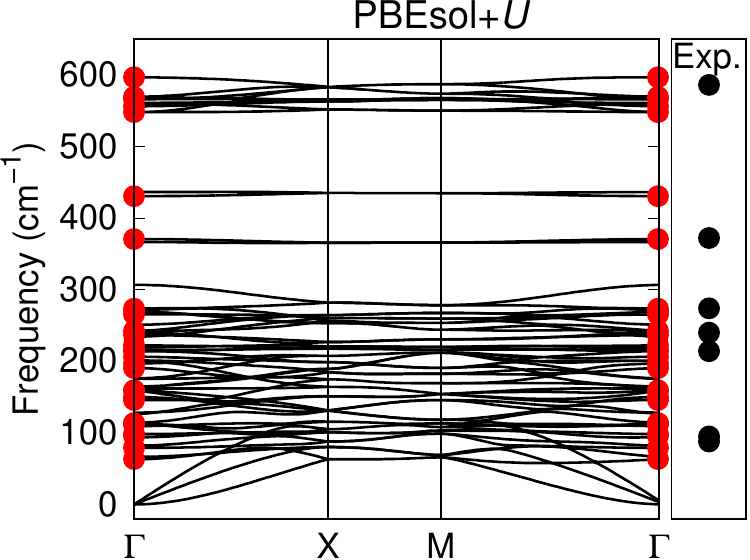}
        \vspace{-4ex}
        \label{fe-ph-2}
      \end{subfigure}
      \hfill
      \begin{subfigure}[t]{.3\textwidth}
        \centering
        \caption{}
        \includegraphics[scale=0.75]{ 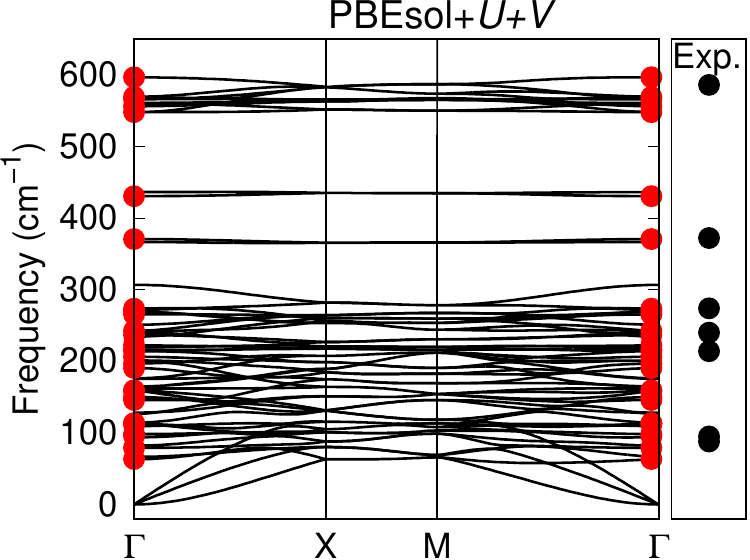}
        \vspace{-4ex}
        \label{fe-ph-3}
      \end{subfigure}
      \newline
      \begin{subfigure}[t]{.32\textwidth}
        \centering
        \caption{}
        \includegraphics[scale=0.75]{ 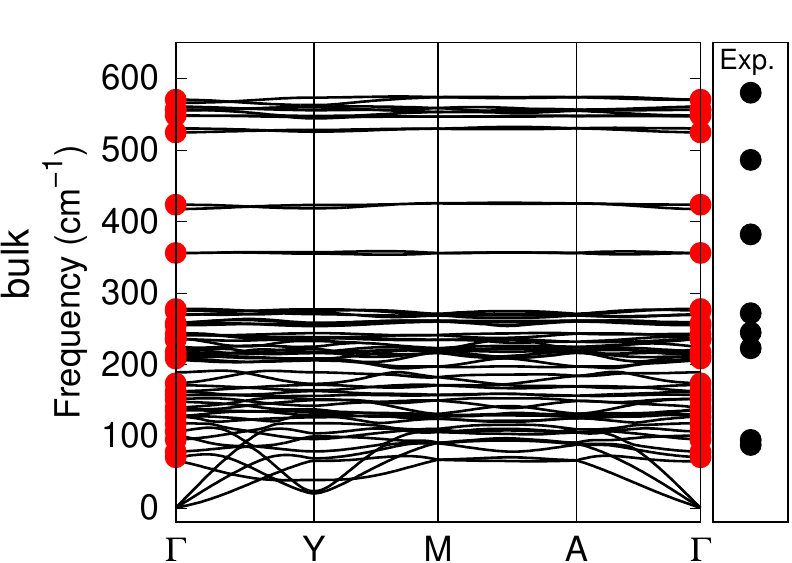}
        \vspace{-4ex}
        \label{fe-ph-4}
      \end{subfigure}
      \hfill
      \begin{subfigure}[t]{.3\textwidth}
        \centering
        \caption{}
        \includegraphics[scale=0.75]{ 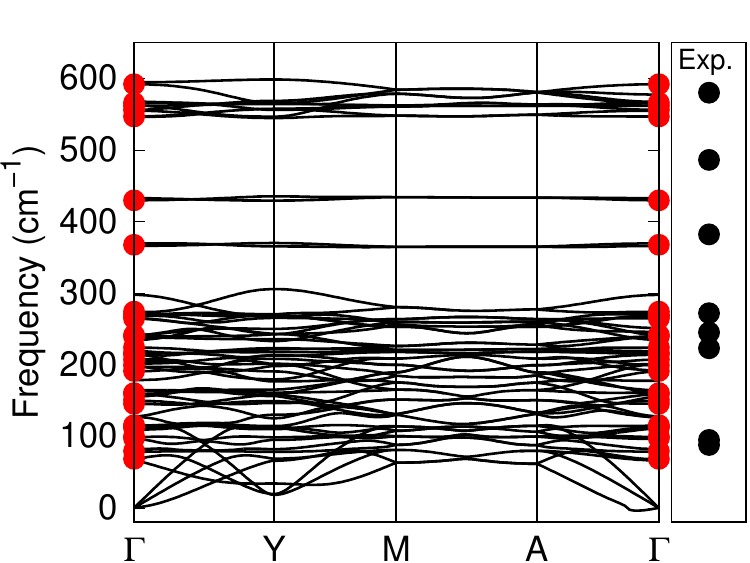}
        \vspace{-4ex}
        \label{fe-ph-5}
      \end{subfigure}
      \hfill
      \begin{subfigure}[t]{.3\textwidth}
        \centering
        \caption{}
        \includegraphics[scale=0.75]{ 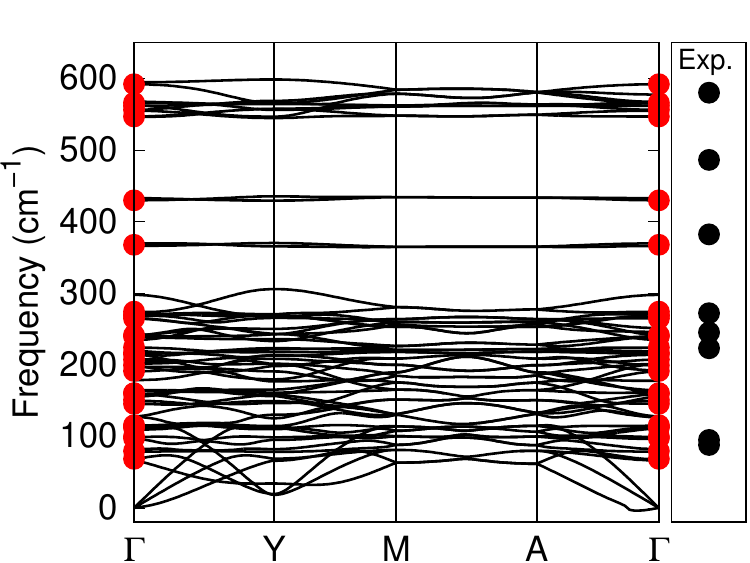}
        \vspace{-4ex}
        \label{fe-ph-6}
      \end{subfigure}
      \captionsetup{justification=Justified}
      \caption{Phonon dispersion for the monolayer (first row) and bulk (second row) of FePS$_3$ as obtained from PBEsol for the distorted structure in c$_i$ symmetry ((a) and (d)), PBEsol+\emph{U} ((b) and (e)) and PBEsol+\emph{U}+\emph{V} ((c) and (f)). The frequencies at the $\Gamma$ point of the Raman active modes are shown by red dots, compared to Raman experiments (black dots).}
      \label{fe-ph-band}
\end{figure*}

The phonon dispersions of FePS$_3$ are presented in \cref{fe-ph-band}, highlighting the Raman active frequencies at the $\Gamma$ point (red dots) to be compared with experimental values extracted from Raman spectra (black dots). The numerical values of selected modes at the $\Gamma$ point are listed in \cref{t6}, and the phonon displacement patterns of all the modes are given in the Supplemental Material. We can see from this table that the Raman peaks do not change significantly between the monolayer and bulk systems~\cite{budniak_spectroscopy_2022}. The calculations provide the full set of phonon frequencies, but experiments only report a few, making the direct comparison in~\cref{fe-ph-band} difficult. There are two sets of vibrations in FePS$_3$: low-frequency phonons (below 200~cm\textsuperscript{-1}) that correspond to vibrations of the heavier Fe ions, and high-frequency phonons associated with the vibrations of P\textsubscript{2}S\textsubscript{6} units~\cite{du_weak_2016,xu_surface_2019,cheng_high-yield_2018,scagliotti_spin_1985,wang_raman_2016,budniak_spectroscopy_2022,martin-perez_direct_2023,liu_direct_2021}. For the low-frequency modes, we cannot compare against experiment (88 and 95~cm$^{-1}$), because we observe several modes below 100~cm\textsuperscript{-1} and we have no information about the experimental mode displacements. For the high-frequency modes, we can infer the mode symmetries by comparison with the well-characterized Raman spectra of lone P\textsubscript{2}S\textsubscript{6} \cite{sourisseau_vibrational_1983} (following the examples of Ref.~\cite{scagliotti_raman_1987} for the bulk and Ref.~\cite{cheng_high-yield_2018} for the quantum sheets).

\begin{figure*}[hbt!]
    \centering
    \begin{subfigure}[t]{.15\textwidth}
        \centering
        \caption{E$_g^1$}
        \vspace{3ex}
        \includegraphics[width=\textwidth]{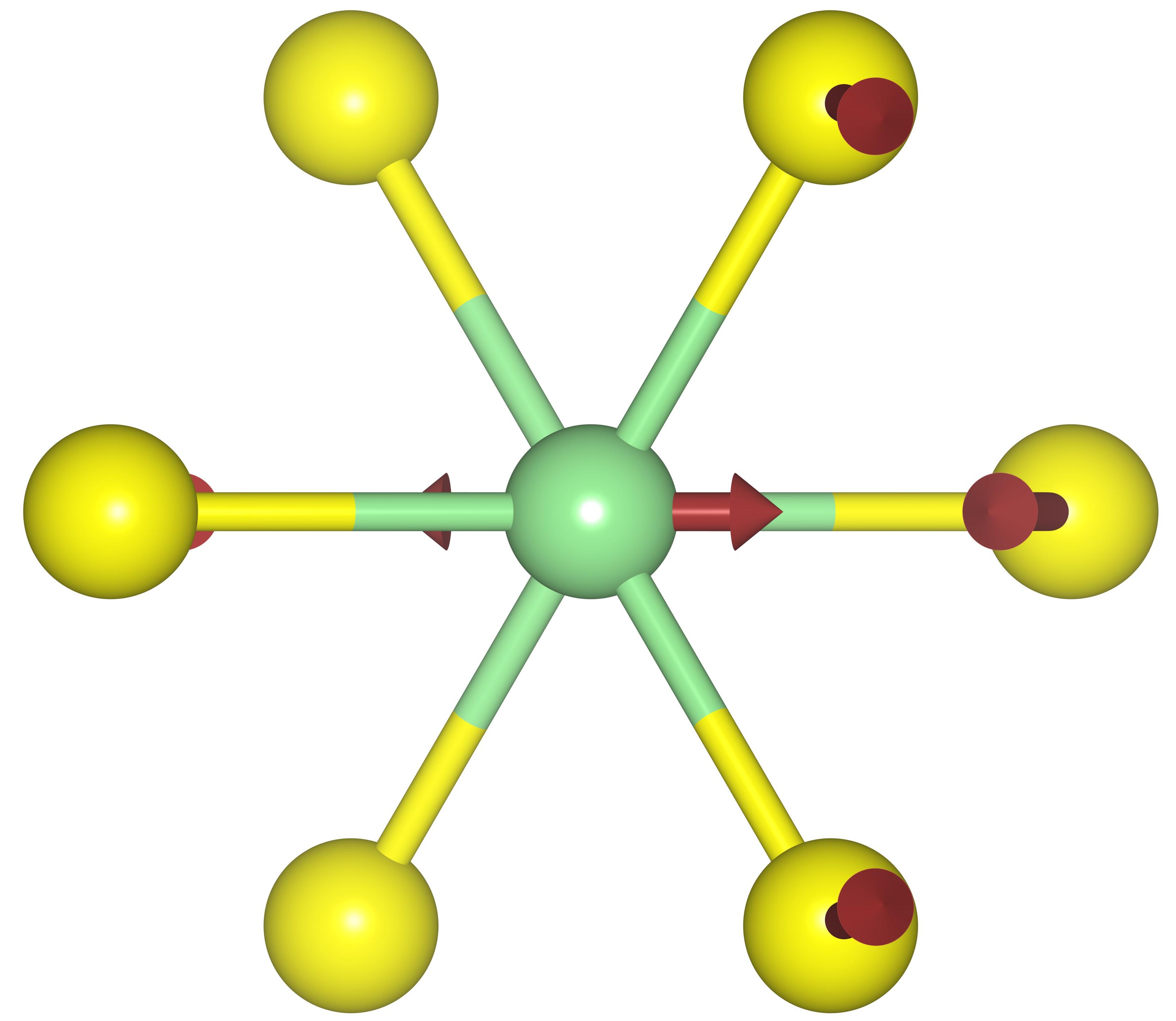}
        \label{eg1}
    \end{subfigure}
    \begin{subfigure}[t]{.15\textwidth}
        \centering
        \caption{A$_{1g}^1$}
        \vspace{3ex}
        \includegraphics[width=\textwidth]{ 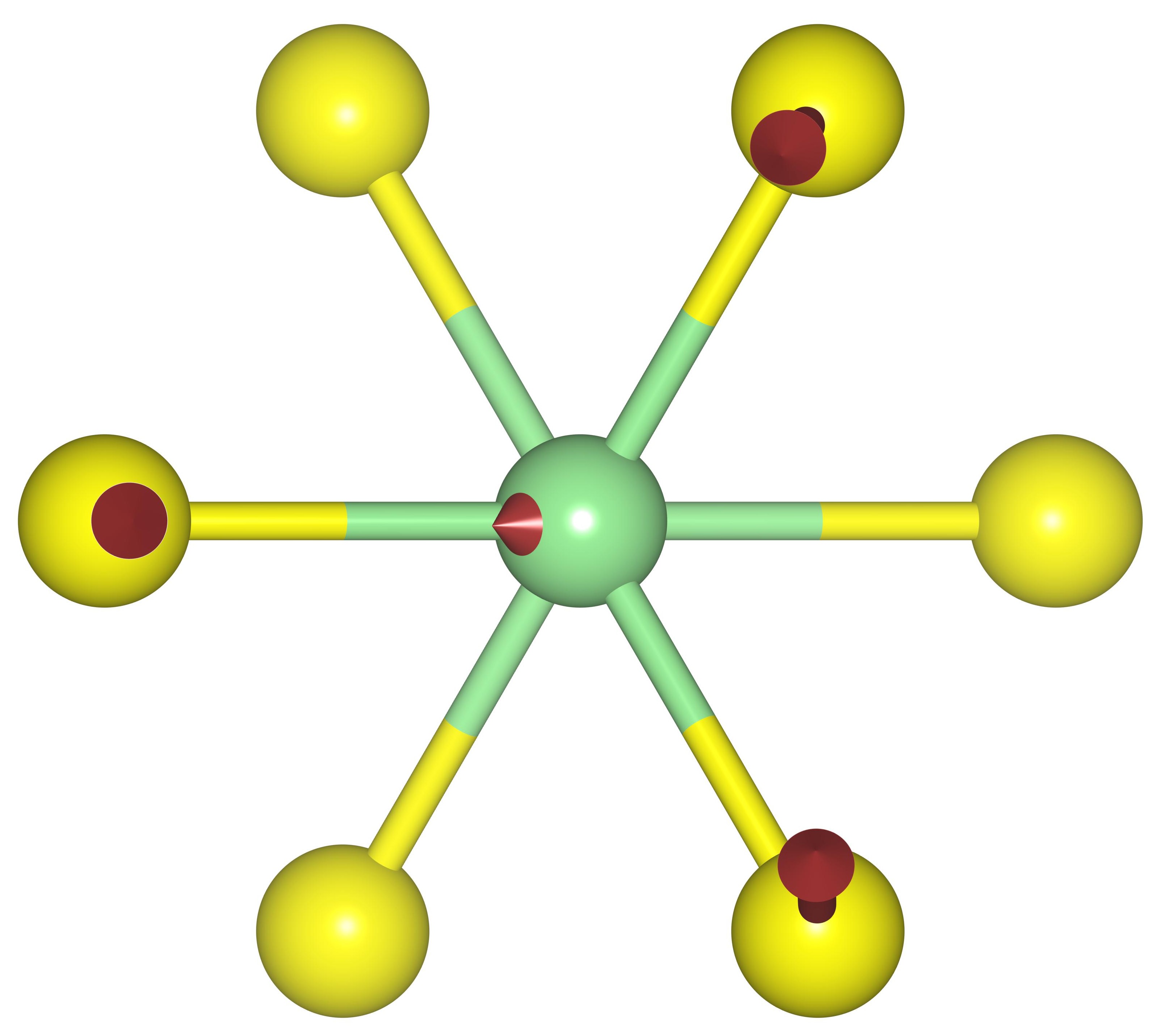}
        \label{a1g1}
    \end{subfigure}
    \begin{subfigure}[t]{.205\textwidth}
        \centering
        \caption{E$_g^2$}
        \vspace{3ex}
        \includegraphics[width=\textwidth]{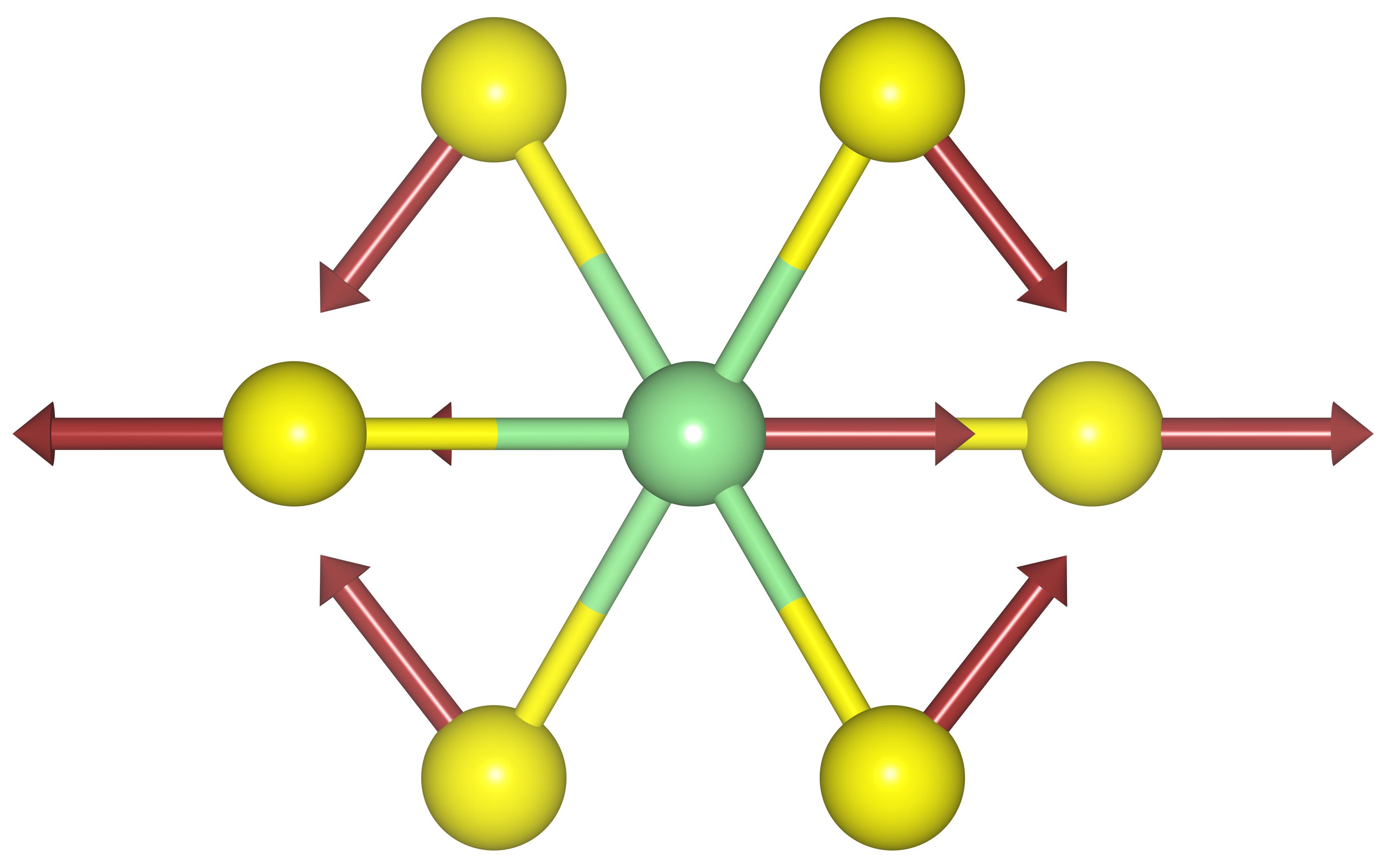}
        \label{eg2}
    \end{subfigure}
        \begin{subfigure}[t]{.205\textwidth}
        \centering
        \caption{A$_{1g}^2$}
        \vspace{0ex}
        \includegraphics[width=\textwidth]{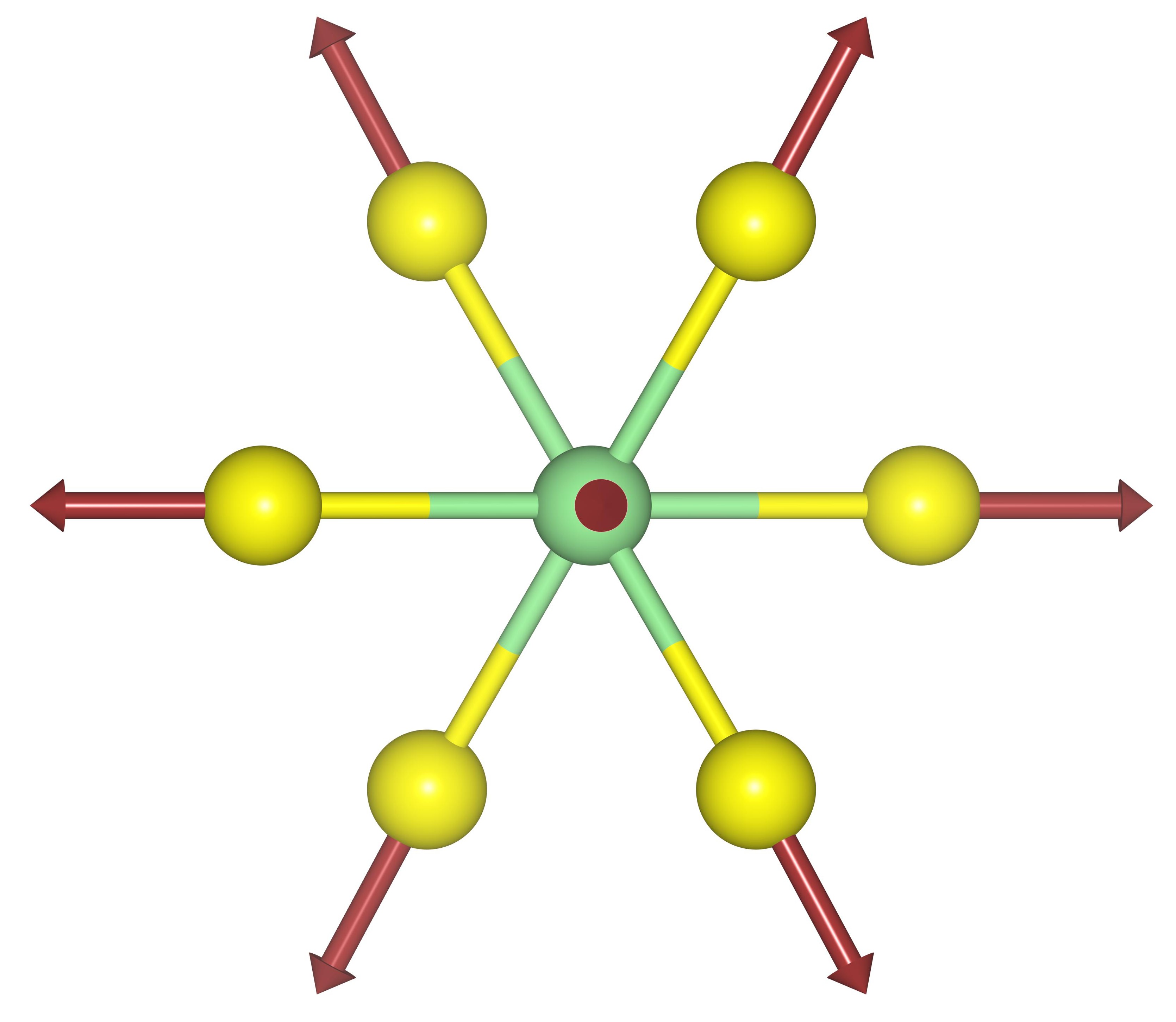}
        \label{a1g2}
    \end{subfigure}
    \begin{subfigure}[t]{.145\textwidth}
         \centering
        \caption{E$_g^3$}
        \vspace{3ex}
        \includegraphics[width=\textwidth]{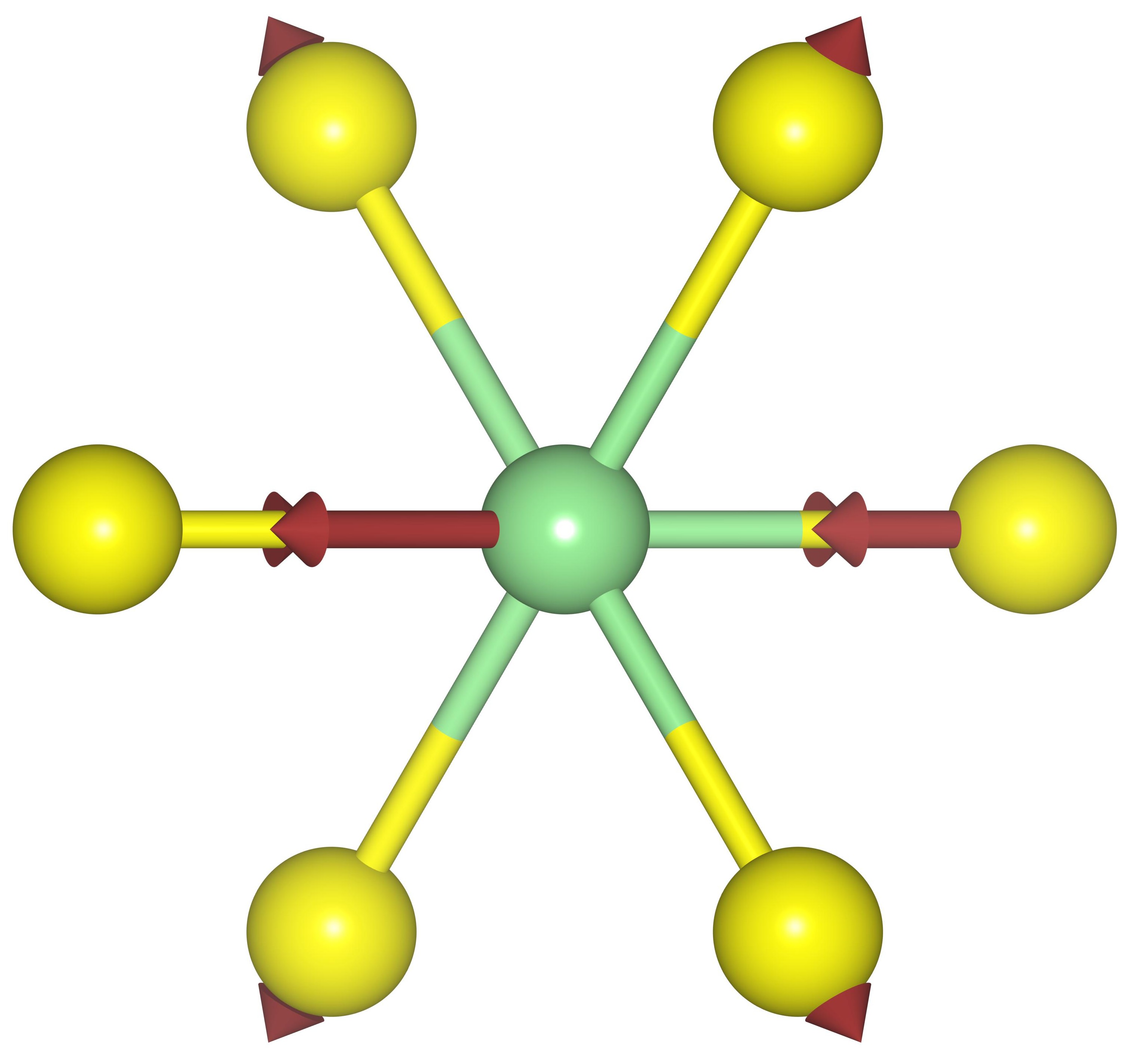}
        \label{eg3}
    \end{subfigure}
    \newline
    \begin{subfigure}[t]{.15\textwidth}
        \centering
        \vspace{-5ex}
        \includegraphics[width=\textwidth]{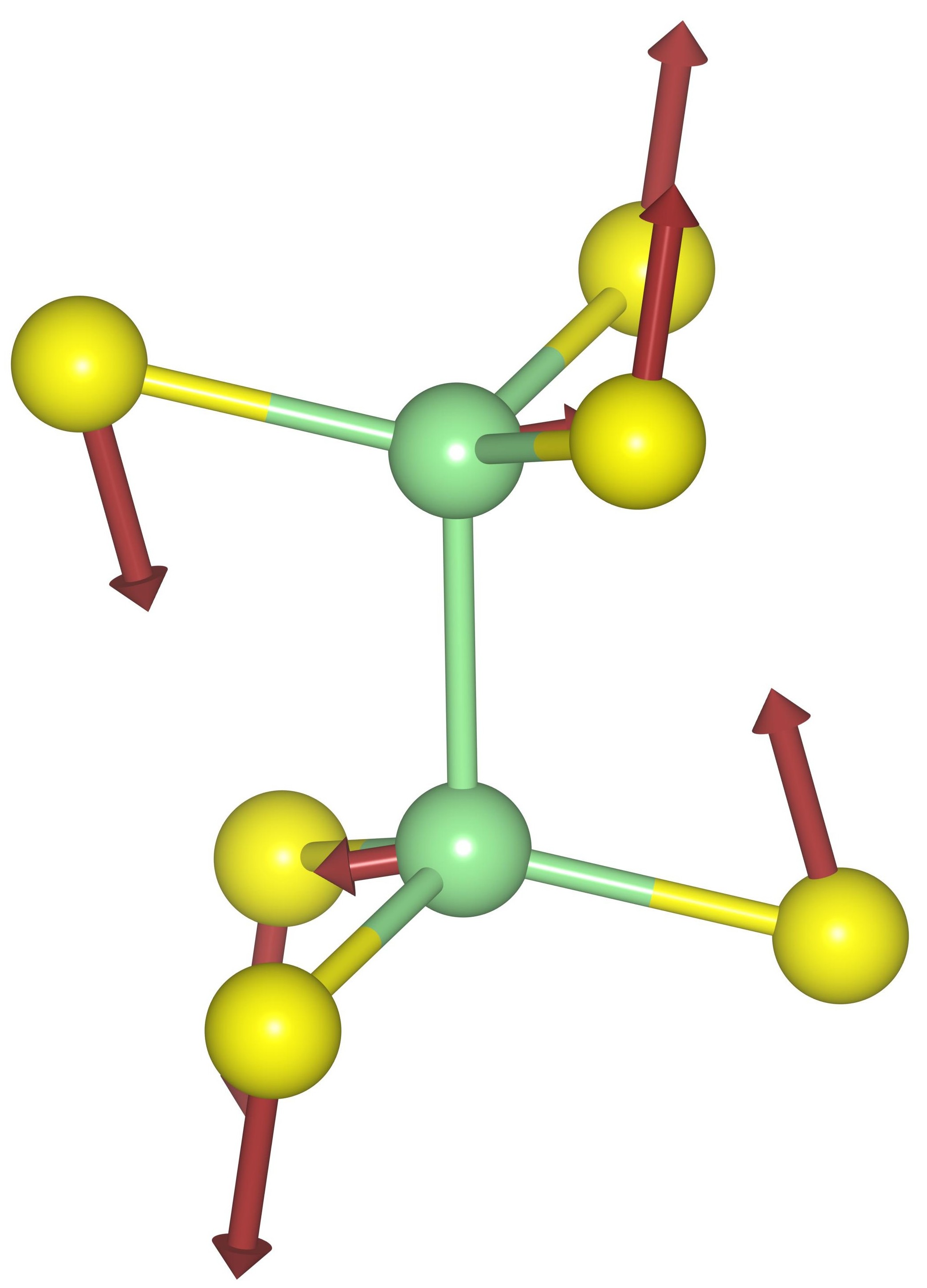}
        \label{eg1-side}
    \end{subfigure}
    \begin{subfigure}[t]{.15\textwidth}
        \centering
        \vspace{-4ex}
        \includegraphics[width=\textwidth]{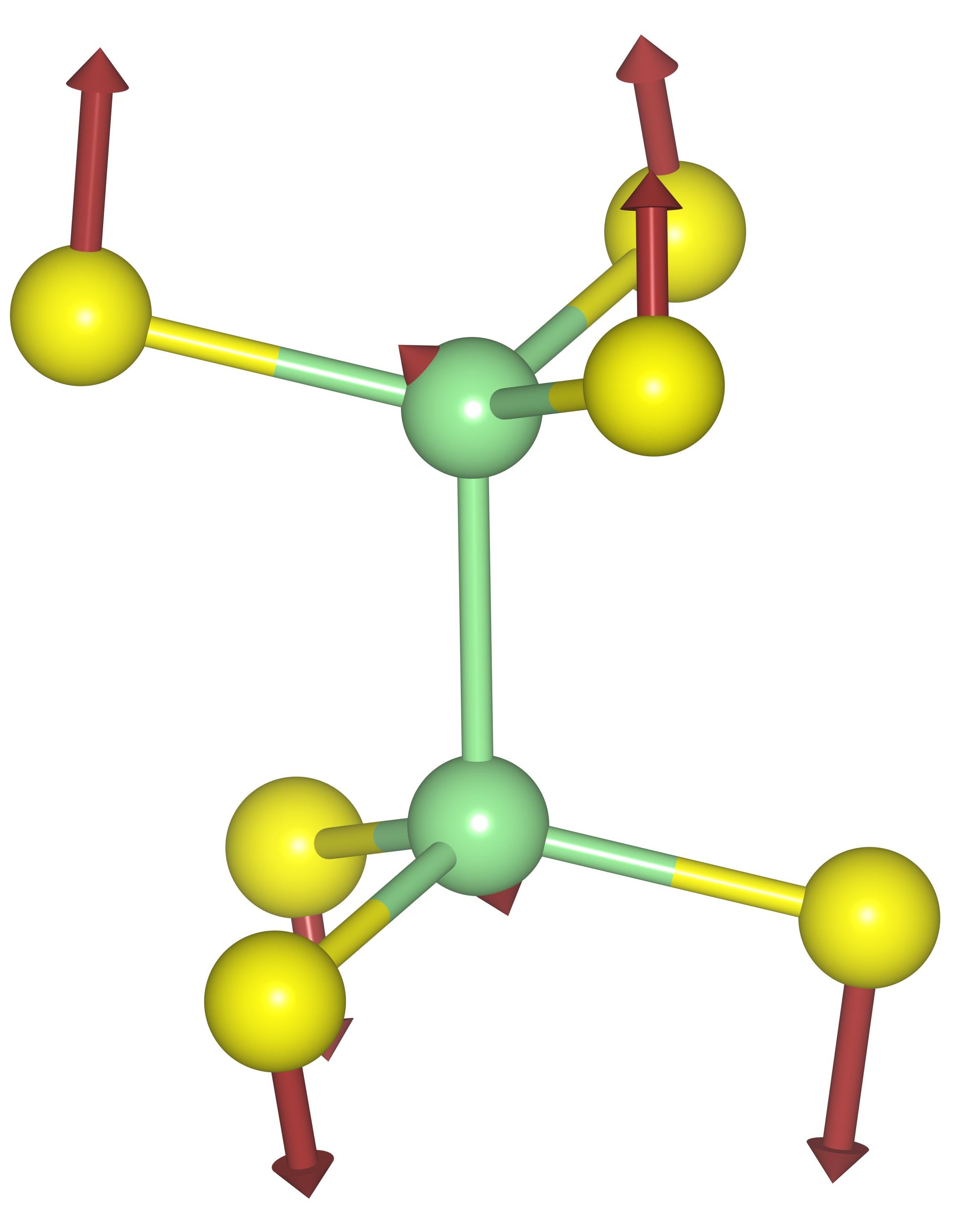}
        \label{a1g1-side}
    \end{subfigure}
    \begin{subfigure}[t]{.22\textwidth}
        \centering
        \vspace{-2ex}
        \includegraphics[width=\textwidth]{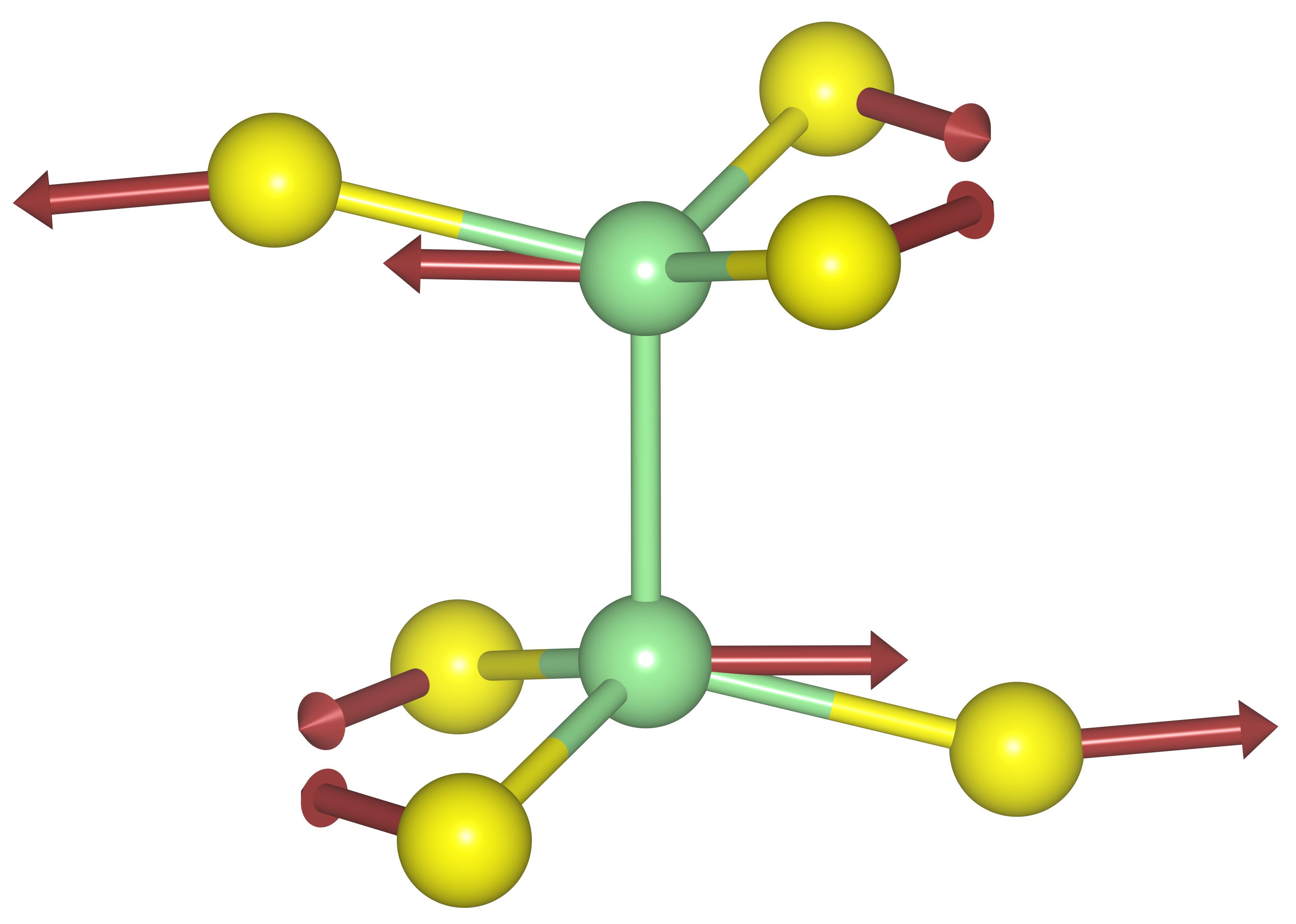}
        \label{eg2-side}
    \end{subfigure}
    \begin{subfigure}[t]{.2\textwidth}
        \centering
        \vspace{-2ex}
        \includegraphics[width=\textwidth]{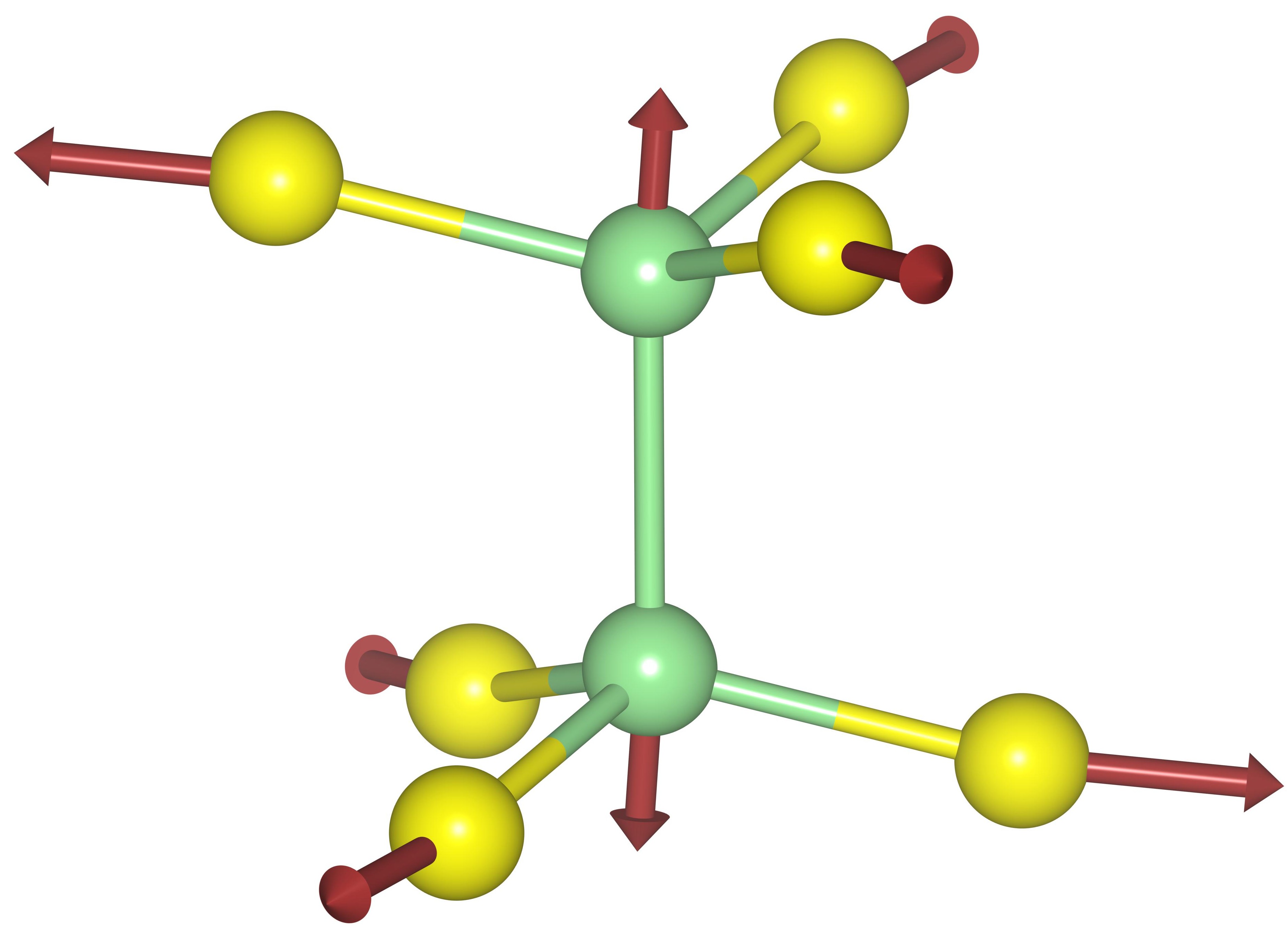}
        \label{a1g2-side}
    \end{subfigure}
    \begin{subfigure}[t]{.15\textwidth}
        \centering
        \vspace{-2ex}
        \includegraphics[width=\textwidth]{ 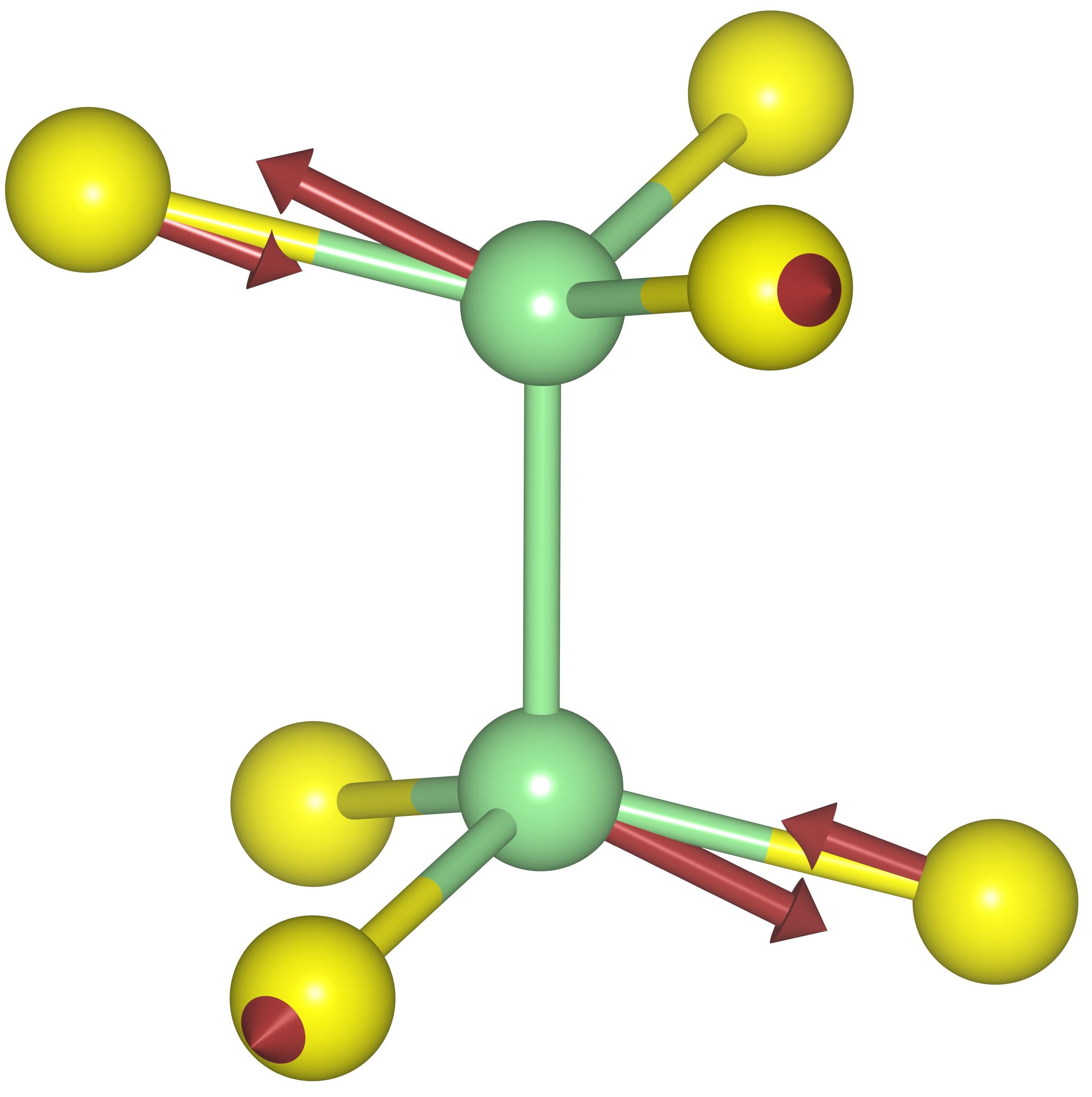}
        \label{eg3-side}
    \end{subfigure}    
    \caption{Top view (upper row) and side view (bottom row) of the displacement patterns for the high-frequency modes of P$_2$S$_6$ units in FePS$_3$ that resemble the vibrations of isolated units, with frequencies listed in \cref{t6}. The analogue mode $A^3_{1g}$ in Refs.~\cite{scagliotti_raman_1987,cheng_high-yield_2018} could not be identified in the calculations.}
    \label{fe-modes}
\end{figure*}

These peaks include three A$_{1g}$ modes, with out-of-plane vibrations of the P$_2$S$_6$ units, and three E$_g$ modes, involving in-plane vibrations and tangential movement of the P--P bond. Such mode displacements are shown in~\cref{fe-modes}; the corresponding frequencies from PBEsol+\emph{U}(+\emph{V}) are reported in~\cref{t6}, in quite good agreement with experiments. We note that due to the crystal environment, the actual mode displacements in FePS$_3$ will differ from those of the molecule shown in Refs.~\cite{scagliotti_raman_1987,cheng_high-yield_2018}. We did not observe an A$_{1g}^3$ mode for bulk FePS\textsubscript{3}, as described in Ref.~\cite{scagliotti_raman_1987}, which identified a low-intensity peak in the experiments around 480~cm\textsuperscript{-1}.

We also note that we could not perform the same comparison for the PBEsol calculations, because in that case the symmetry of the system was reduced. The one exception to this is the $A^2_{1g}$ peak, to which PBEsol assigns a frequency of 358~cm\textsuperscript{-1} -- in worse agreement with experiments (380~cm\textsuperscript{-1}) than PBEsol+\emph{U}(+\emph{V}) calculations (around 370~cm$^{-1}$).

\begin{table*}[t]
    \caption{\label{t6}  The Raman active modes for FePS$_3$ from calculations and experiments for bulk and monolayer. The experimental data is for 4~K in the bulk and 273~K for the monolayer. In Ref.~\cite{scagliotti_raman_1987} it is shown that the Raman data for the bulk does not significantly differ between room temperature and liquid Helium temperature; we assume here that the same behavior holds for the 2D system, and compare the room temperature Raman data from Ref.~\cite{cheng_high-yield_2018} with our first-principles results. Note that the modes are labeled according to $D3d$ symmetry of P$_2$S$_6$ molecule following Refs.~\onlinecite{scagliotti_raman_1987,cheng_high-yield_2018}. In 
    FePS$_3$ crystals the E$_g$ modes are split into A$_g$ and B$_g$ modes and what we show here is the frequency of the A$_g$ mode. }
    \centering
    \renewcommand{\arraystretch}{1.5}
    \begin{tabularx}{\textwidth}{ X X D..{8.5} *{2}{D..{8.4}} *{3}{D..{8.5}}}
    \hline
    \hline
    & 
    & \multicolumn{1}{c}{E$_g^1$ }
    & \multicolumn{1}{c}{A$_{1g}^1$ }
    & \multicolumn{1}{c}{E$_g^2$ }
    & \multicolumn{1}{c}{A$_{1g}^2$ }
    & \multicolumn{1}{c}{E$_g^3$}  \Tstrut \\[0.1cm]
    \hline
    \multirow{3}{*}{bulk}
    & Expt.                         & 223 & 245 & 272 & 382   & 580  \\
    & PBEsol+\emph{U}               & 224 & 240 & 269 & 370   & 547   \\
    & PBEsol+\emph{U}+\emph{V}      & 224 & 240 & 269 & 369.5 & 546.5  \\
    \hline
    \multirow{3}{*}{monolayer}
    & Expt.                         & 214 & 240  & 274  & 372  & 586 \\
    & PBEsol+\emph{U}               & 222 & 237  & 269  & 371  & 548 \\
    & PBEsol+\emph{U}+\emph{V}      & 222 & 237  & 269  & 371  & 548 \\
    \hline
    \hline
    \end{tabularx}
\end{table*}

Finally, the bulk phonon frequencies are calculated including the non-analytic terms using Born effective charges (BECs) and the dielectric tensor, $\epsilon^{\infty}$, which are reported in \cref{t10} and to our knowledge have not been reported yet for FePS$_3$. Since the effects of Hubbard \emph{V} parameters on vibrational properties are small, we use the BECs and $\epsilon^{\infty}$ from PBEsol+\emph{U} for PBEsol+\emph{U}+\emph{V} phonons. The different symmetries of the PBEsol and PBEsol+\emph{U} ground-state structures are reflected in the different form of the tensors, with the appearance of additional symmetry-enforced vanishing values in the more symmetric case with Hubbard corrections, and the presence of more inequivalent atoms in the PBEsol case with reduced symmetry. With both functionals, the anisotropic character of the dielectric tensor --- with different values along the $x$ and $y$ directions --- is consistent with the monoclinic distortion of the layers, which is larger in PBEsol calculations and is associated with the zigzag  spin configuration.  The large numerical values in the PBEsol dielectric tensor are attributed to the fact that the system has a small band gap in this case (see \cref{fe-band-4}).\\
Last, we stress that the small negative frequencies of the long-wavelength acoustic branch along the $\Gamma$-A path for bulk FePS$_3$ are not a sign of physical instability, but most likely result from  an insufficient supercell size along the $c$ axis in the calculation of phonon frequencies by finite differences. Indeed, the large number of atoms in the primitive cell limits the extension of the current $2 \times 2\times 2$ supercell owing to the associated  computational cost. Since the $\Gamma$-A line corresponds to the vertical (i.e.\ orthogonal to the layers) direction, increasing the supercell size along the vertical  $c$ axis might solve this minor issue, especially when combined with the correction schemes of Ref.~\onlinecite{lin_general_2022} that also require a sufficiently large supercell, but lies beyond the scopes of the present study.

%-------------------------------------------------------------------------------------

\begin{table}[h]
\caption{\label{t10} The dielectric tensor and BECs of symmetrically-inequivalent atoms in bulk FePS$_3$, as given by PBEsol and PBEsol+\emph{U} calculations in the Cartesian framework. Note that the symmetry of the system is lower in the case of PBEsol resulting in more inequivalent atoms.}
\centering
\renewcommand{\arraystretch}{1.3}
\begin{tabularx}{\columnwidth}{c c c }
\hline
\hline
 &  PBEsol  &  PBEsol+\emph{U} \\ [0.1cm]
\hline 
   $\epsilon^{\infty}$ &
   $\begin{pmatrix}  
   20.15 & 0.34 & -2.08\\
   0.34 & 16.68 & -0.51 \\
   -2.08 & -0.51 & 7.01 \end{pmatrix}$
   & 
   $\begin{pmatrix}  
   7.67 & 0.0 & -0.54\\
   0.0 & 7.89 & 0.0 \\
   -0.54 & 0.0 & 4.40  \end{pmatrix}$
   \\
   \hline
 Z$_{Fe}$ &  
 $\begin{pmatrix} 
 1.28 & 0.51 & -0.12\\
 0.20 & 1.50 & -0.40 \\
 -0.29 & -0.49 & 0.80 \end{pmatrix}$ 
 & 
 $\begin{pmatrix} 
 1.92 & 0.0 & -0.28\\
 0.0 & 2.44 & 0.0 \\
 -0.17 & 0.0 & 1.02  \end{pmatrix}$ \\ [0.1cm]
\hline
  Z$_{P}$ &  
  $\begin{pmatrix}  
   3.08 &  0.20 & -0.44\\
   0.12 &  3.34 & -0.40 \\
  -0.46 & -0.62 &  0.51 \end{pmatrix}$ 
  & 
   $\begin{pmatrix} 
  3.11 & 0.0 & -0.33\\
  0.0 & 3.04 & 0.0 \\
  -0.39 & 0.0 & 0.72 \end{pmatrix}$ \\ [0.1cm]
 \hline
  Z$_{S_1}$ &  
  $\begin{pmatrix}  
   -1.87 &  -0.40 & 0.62\\
   -0.16 &  -1.40 & 0.31 \\
  0.30 & 0.571 &  -0.62 \end{pmatrix}$  
  & 
   $\begin{pmatrix} 
  -2.18 & 0.0 & 0.51\\
  0.0 & -1.42 & 0.0 \\
  0.25 & 0.0 & -0.65 \end{pmatrix}$ \\ [0.1cm]
 \hline
   Z$_{S_2}$ &  
  $\begin{pmatrix}  
   -1.41 &  -0.22 & 0.20\\
   -0.01 &  -1.77 & 0.03 \\
  -0.21 & 0.07 &  -0.38 \end{pmatrix}$ 
  & 
  $\begin{pmatrix}  
   -1.43 &  -0.12 & 0.05\\
   0.11 &  -2.03 & -0.47 \\
  0.16 & -0.46 &  -0.54 \end{pmatrix}$ \\ [0.1cm]
 % \cline{1-2}
 \hline
    Z$_{S_3}$ & 
  $\begin{pmatrix}  
   -1.09 &  -0.09 & 0.15\\
   -0.15 &  -1.66 & 0.47 \\
  0.23 & 0.47 &  -0.30 \end{pmatrix}$ 
  & 
  ---
 \\
\hline
\hline
\end{tabularx}
\end{table}

\subsection{CrI$_3$}
\label{subsec2}

\begin{figure}[t]
\includegraphics[width=1\columnwidth]{ 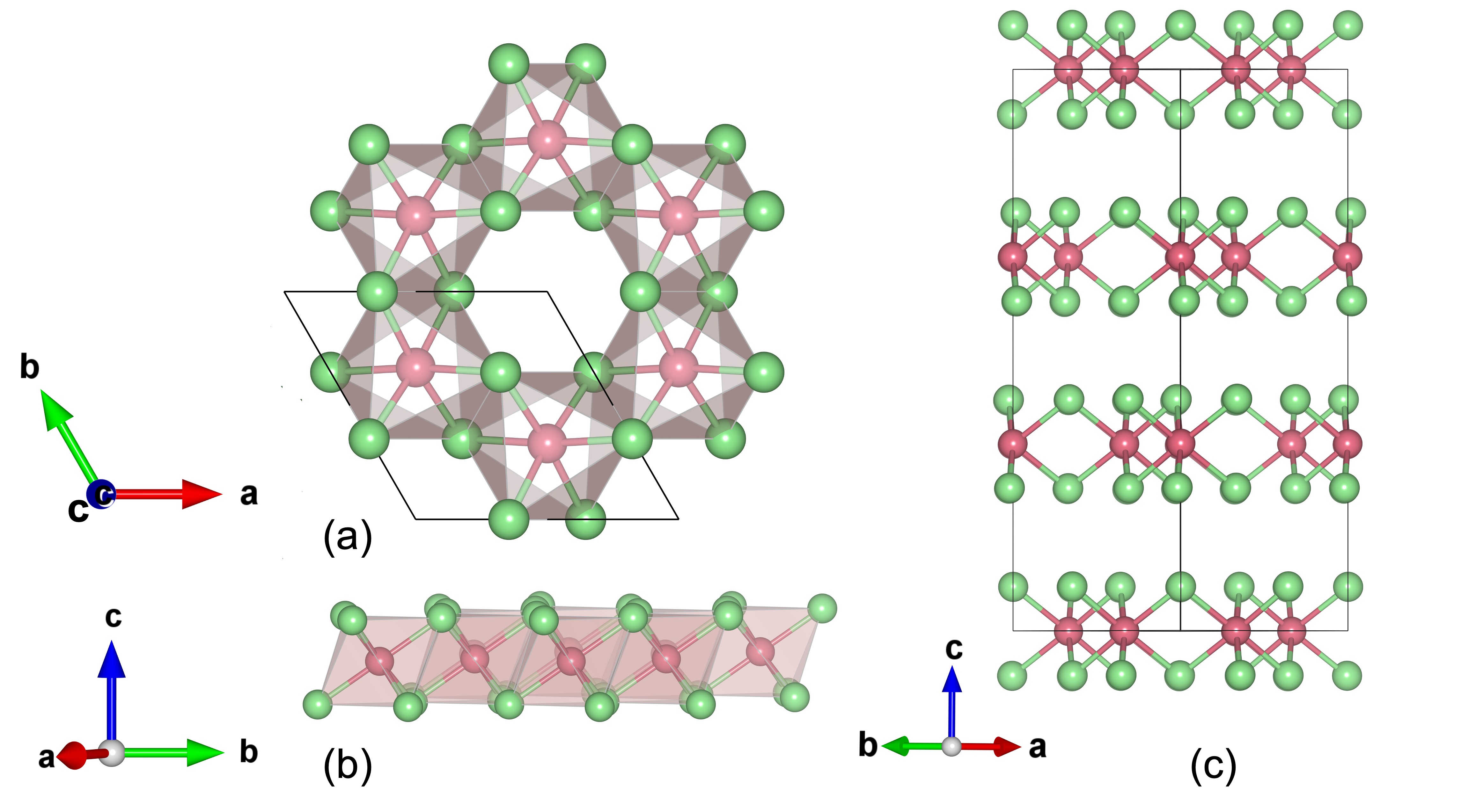}
\caption{(a) Top view and (b) side view of monolayer CrI$_3$. The unit cell is shown with black lines. The pink and green balls correspond to Cr and I atoms respectively. (c) The bulk system is composed of monolayers stacked along the $c$ axis.}
\label{fig3}
\end{figure}

Bulk CrI$_3$ is a vdW ferromagnet with monoclinic AlCl$_3$ structure (space group C2/m) at high temperatures and rhombohedral BiI$_3$ structure (space group $R\Bar{3}$) at low temperatures. 
In both phases, the Cr ions form a honeycomb lattice sandwiched between two layers of iodine (\cref{fig3}). The primitive unit cell contains 8 atoms including two Cr and six I atoms. Each Cr$^{3+}$ ion has an electronic configuration of $3d^{3}4s^{0}$ and six nearest neighbors I$^{-}$ forming edge-sharing octahedra.

The structural properties and magnetization of monolayer and bulk CrI$_3$ are reported in \cref{t2}, with small variations between the two systems. For the lattice constant ($a$) and Cr-I distance ($l_{Cr-I}$), the results from PBEsol exhibit closer agreement with experiments compared to PBEsol+$U$($+V$). In the bulk, the maximum discrepancy between the calculated values for the interlayer spacing between Cr atoms ($d$) and the experimental data is less than 1$\%$, with the PBEsol+\emph{U}+\emph{V} exhibiting the least agreement. The last line in \cref{t2} shows the results when a different value of $U$ is computed separately for spin-up and spin-down channels to address the challenge of the shift of the spin-minority conduction bands as a result of Hubbard corrections, which will be extensively explained later. We only note here that the structural properties and magnetization of monolayer CrI$_3$ from PBEsol and PBEsol+$U_\uparrow$+$U_\downarrow$ are very similar. Turning to the magnetization ($m_{Cr}$), Cr$^{3+}$ ions with out-of-plane magnetic anisotropy are expected to have a nominal spin of $S=3/2$~\cite{seyler_ligand-field_2018,richter_temperature-dependent_2018}. The magnetic moment from PBEsol is more consistent with this picture, while that from PBEsol+$U$(+$V$) is larger with respect to the experiments and more consistent with the results from high-accuracy quantum Monte Carlo calculations~(about 3.62$\mu_B$)~\cite{staros_combined_2022}. PBEsol+\emph{U}+\emph{V} predicts a slightly smaller magnetization than PBEsol+$U$ because of the delocalization effects induced by the Hubbard \emph{V}.

\begin{table}[t]
    \caption{\label{t2} In-plane lattice constant $a$ (in \AA),  distance $l_{\rm Cr-I}$ between Cr and nearest I atoms (in \AA), layer spacing $d$ between Cr layers (in \AA) and magnetization $m_{Cr}$ of Cr atoms (in $\mu_B$, calculated from $m^I = \sum_{m} \left( n^{I\uparrow}_{m m} - n^{I\downarrow}_{m m} \right)$ as described in section \ref{sec2}), computed with different functionals for monolayer and bulk CrI$_{3}$. Experimental values are also reported~\cite{mcguire_coupling_2015,dillon_magnetization_1965}.}
    \centering
    \renewcommand{\arraystretch}{1.5}
    \begin{tabularx}{\columnwidth}{c X *{3}{D..{2.4}} D..{2.3}}
    \hline
    \hline
    &  & \multicolumn{1}{c}{$a$  } & \multicolumn{1}{c}{$l_{\rm Cr-I}$ } & \multicolumn{1}{c}{$d$ } & \multicolumn{1}{c}{$m_{Cr}$}  \Tstrut \\[0.1cm]
    \hline
    \parbox[t]{1.2em}{\multirow{4}{*}{\rotatebox[origin=c]{90}{bulk}}}& Expt. & 6.867 & 2.727 & 6.602 & 3.10 \\
    & PBEsol                    & 6.823 & 2.695 & 6.578 & 3.19 \\
    & PBEsol+\emph{U}           & 6.999 & 2.783 & 6.614 & 3.99\\
    & PBEsol+\emph{U}+\emph{V}  & 6.988 & 2.776 & 6.649 & 3.93\\
    \hline
    \parbox[t]{1.2em}{\multirow{4}{*}{\rotatebox[origin=c]{90}{monolayer}}}
    & PBEsol & 6.817  & 2.692 & \multicolumn{1}{c}{---}  & 3.18 \\
    & PBEsol+\emph{U}                    & 6.978   & 2.774 & \multicolumn{1}{c}{---} & 3.97\\
    & PBEsol+\emph{U}+\emph{V}           & 6.971   & 2.767 & \multicolumn{1}{c}{---} & 3.89\\
    & PBEsol+$U_\uparrow$+$U_\downarrow$ & 6.826   & 2.699 & \multicolumn{1}{c}{---} & 3.21\\
    \hline
    \hline
    \end{tabularx}
\end{table}

\begin{figure*}[h]
    \captionsetup{singlelinecheck = false, justification=raggedright}
    \centering
    \begin{subfigure}[t]{.48\textwidth}
        \centering
        \caption{}
        \includegraphics[width=\textwidth]{ 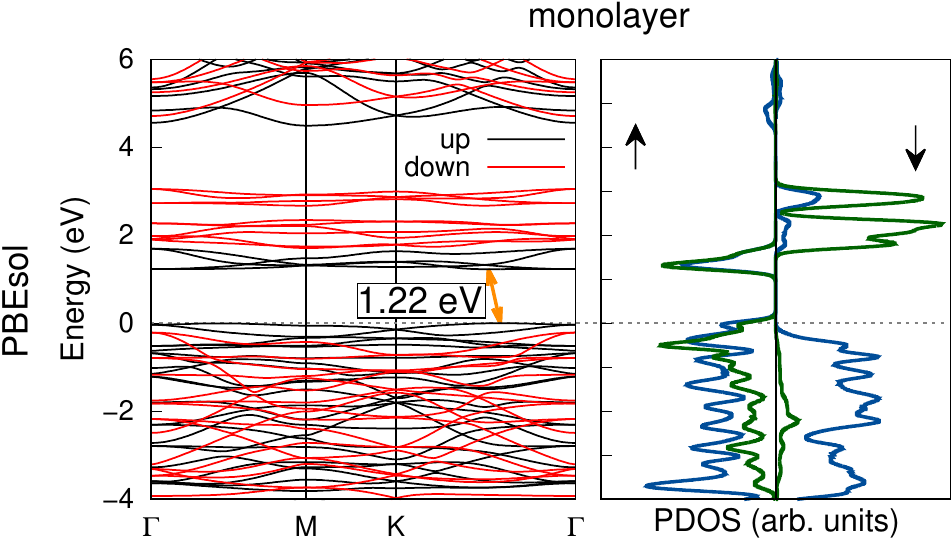}
        \vspace{-4ex}
        \label{cr-band-1}
    \end{subfigure}
    \hfill
    \begin{subfigure}[t]{.46\textwidth}
        \centering
        \caption{}
        \includegraphics[width=\textwidth]{ 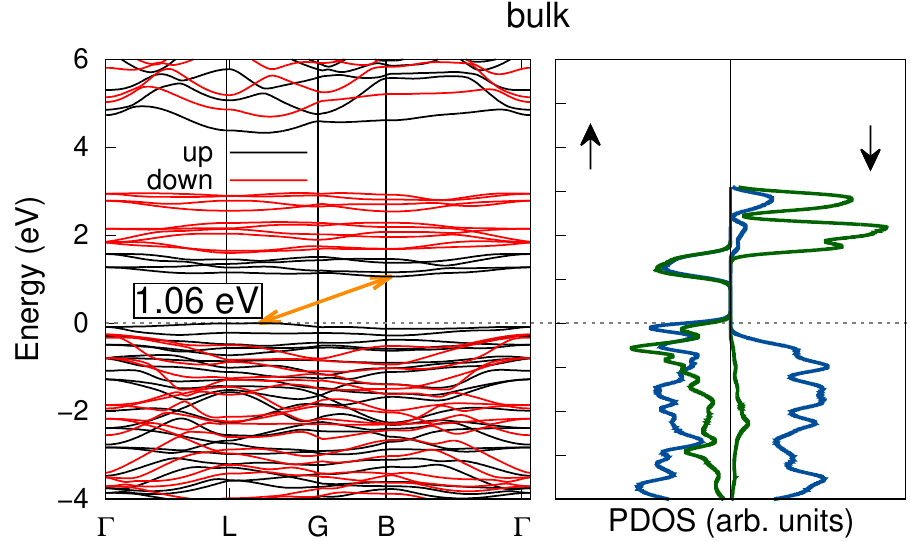}
        \vspace{-4ex}
        \label{cr-band-2}
    \end{subfigure}
    \newline
    \begin{subfigure}[t]{.48\textwidth}
        \centering
        \caption{}
        \includegraphics[width=\textwidth]{ 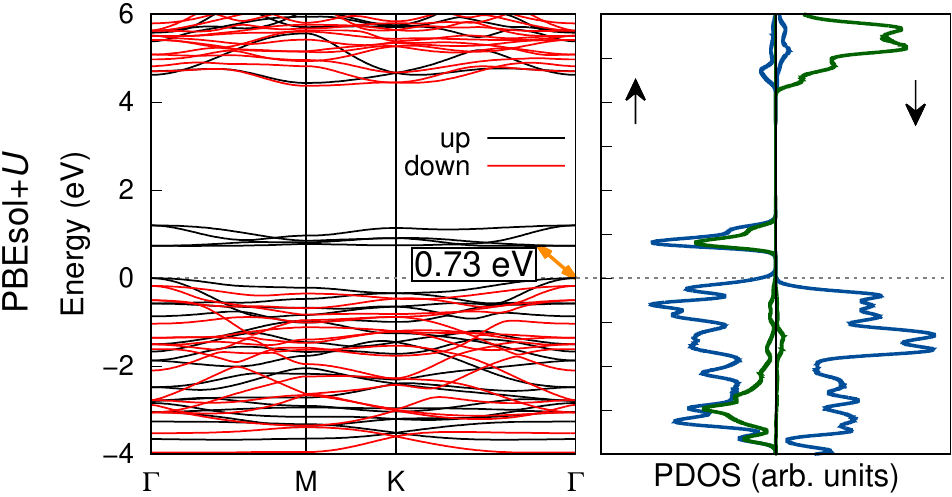}
        \vspace{-4ex}
        \label{cr-band-3}
    \end{subfigure}
    \hfill
    \begin{subfigure}[t]{.46\textwidth}
        \centering
        \caption{}
        \includegraphics[width=\textwidth]{ 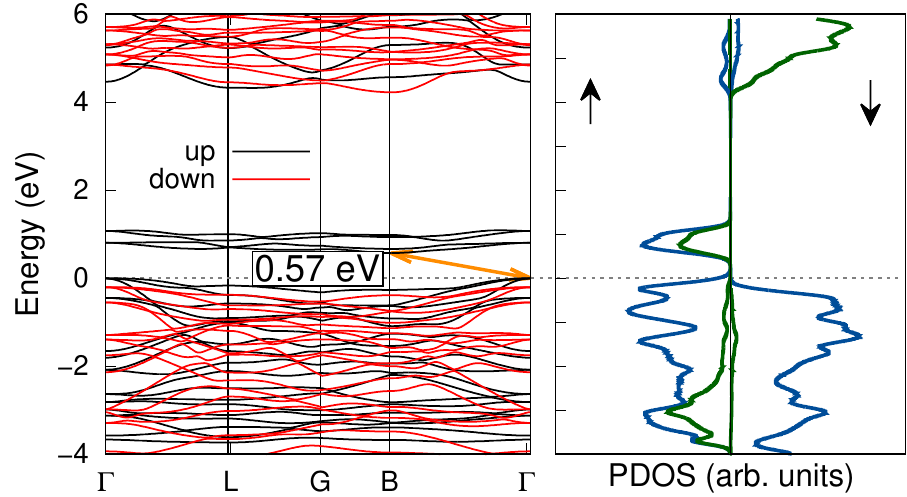}
        \vspace{-4ex}
        \label{cr-band-4}
    \end{subfigure}
    \newline
    \begin{subfigure}[t]{.48\textwidth}
        \centering
        \caption{}
        \includegraphics[width=\textwidth]{ 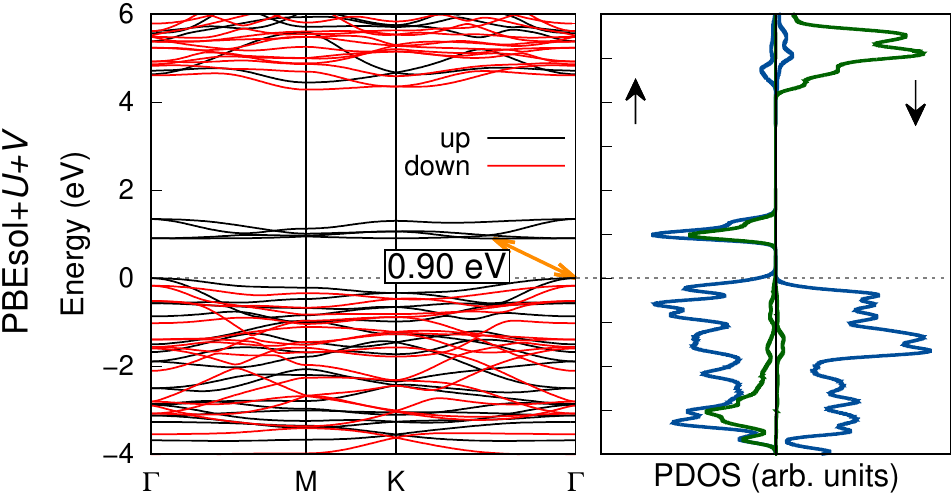}
        \vspace{-4ex}
        \label{cr-band-5}
    \end{subfigure}
    \hfill
    \begin{subfigure}[t]{.46\textwidth}
        \centering
        \caption{}
        \includegraphics[width=\textwidth]{ 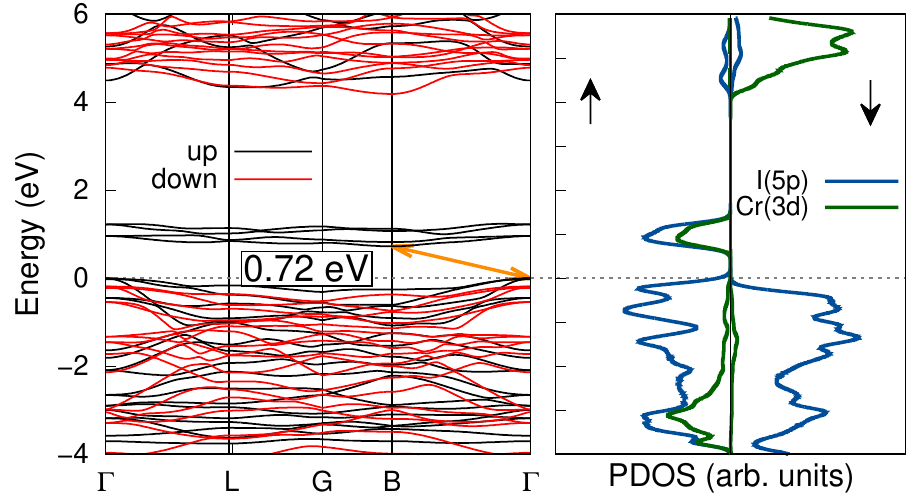}
        \vspace{-4ex}
        \label{cr-band-6}
    \end{subfigure}
    \captionsetup{justification=Justified}
    \caption{The bandstructure and PDOS of CrI$_3$ monolayer (first column) and bulk (second column) from (a) and (b) PBEsol, (c) and (d) PBEsol+\emph{U} and (e) and (f) PBEsol+\emph{U}+\emph{V}. The band gap values are given in the figure.}
    \label{cr-band}
\end{figure*}

The band structures and PDOS of CrI$_3$ using PBEsol and PBEsol+\emph{U}(+\emph{V}) calculations are shown in \cref{cr-band}. The valence band maximum (VBM) is at the $\Gamma$ point in the PBEsol+\emph{U}(+\emph{V}) cases, in agreement with angle-resolved photoemission spectroscopy (ARPES)~\cite{kundu_valence_2020}, and as also captured by more expensive extended quasiparticle self-consistent GW (QSG$\widehat{W}$) calculations~\cite{acharya_electronic_2021}. Bulk CrI$_3$ has an experimental optical band gap of 1.2~eV~\cite{dillon_magnetization_1965}; this serves as a lower bound to the quasiparticle gap that GW calculations at different levels of self-consistency predict to be between 2.2 and 3.25 eV~\cite{lee_role_2020,molina-sanchez_magneto-optical_2020,acharya_real-_2022} owing to a significant exciton binding energy.

  \begin{figure*}[h]
     \centering
     \begin{subfigure}[t]{.46\textwidth}
         \centering
         \caption{$U$=6.54 eV}
         \includegraphics[width=\textwidth]{  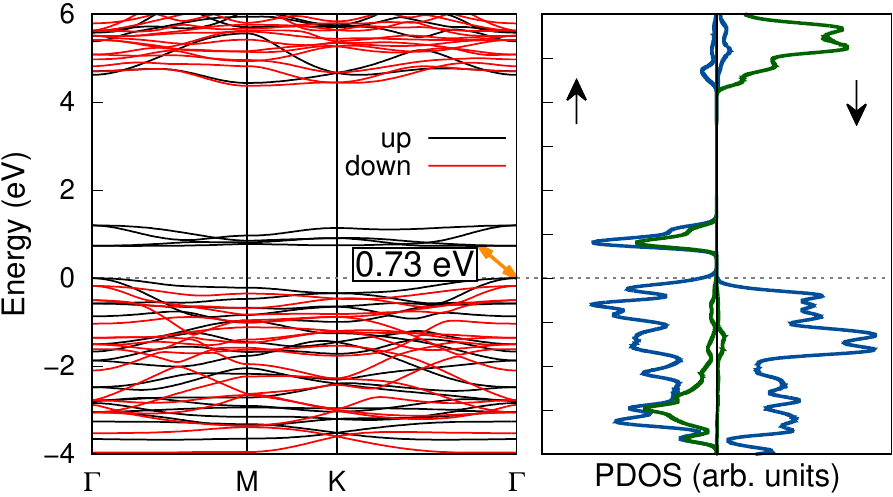}
         \label{cr-u}
     \end{subfigure}
     \hfill
     \begin{subfigure}[t]{.46\textwidth}
         \centering
         \caption{$U_\uparrow$= 1.72 eV,  $U_\downarrow$=0.31 eV}
         \includegraphics[width=\textwidth]{  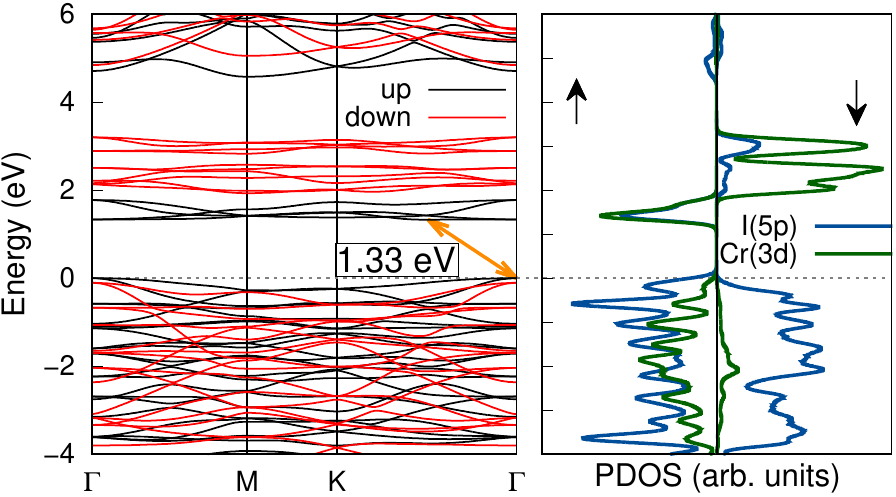}
         \label{spin_res-u}
     \end{subfigure}
     \caption{The band structure and PDOS of CrI$_3$ monolayer  from (a) PBEsol+\emph{U} and (b) PBEsol+$U_\uparrow$+$U_\downarrow$.}
     \label{spin-res-band}
\end{figure*}

\begin{figure*}[ht!]
    \captionsetup{singlelinecheck = false, justification=raggedright}
    \centering
    \begin{subfigure}[t]{.33\textwidth}
             \centering
             \caption{}
             \includegraphics[scale=0.75]{ 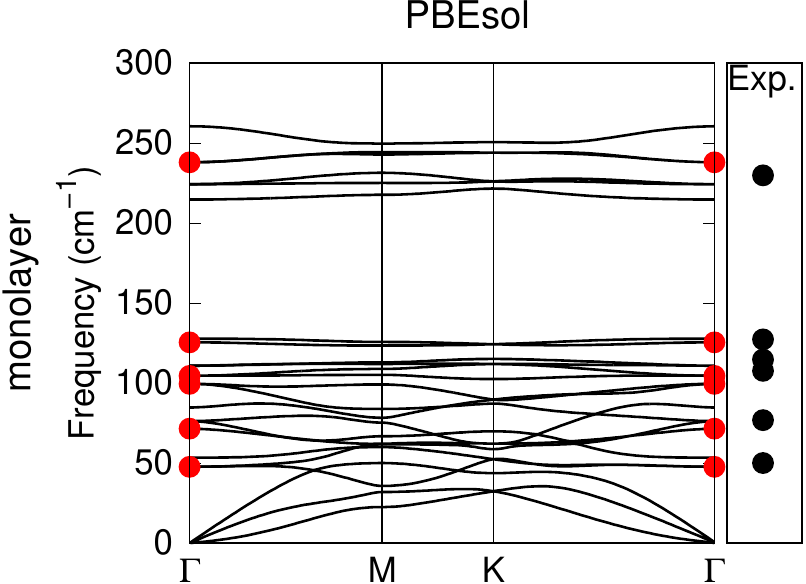}
             \vspace{-4ex}
             \label{cr-ph-1}
    \end{subfigure}
    \hfill
    \begin{subfigure}[t]{.3\textwidth}
             \centering
             \caption{}
             \includegraphics[scale=0.75]{ 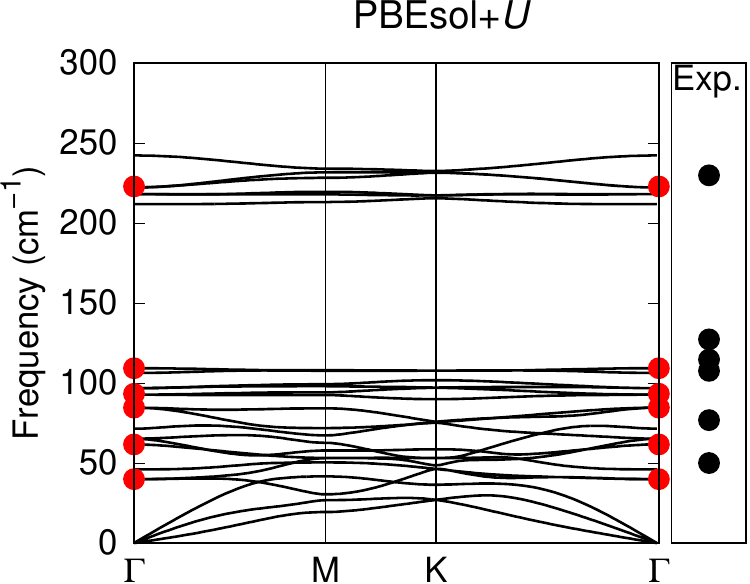}
             \vspace{-4ex}
             \label{cr-ph-2}
    \end{subfigure}
    \hfill
    \begin{subfigure}[t]{.3\textwidth}
             \centering
             \caption{}
             \includegraphics[scale=0.75]{  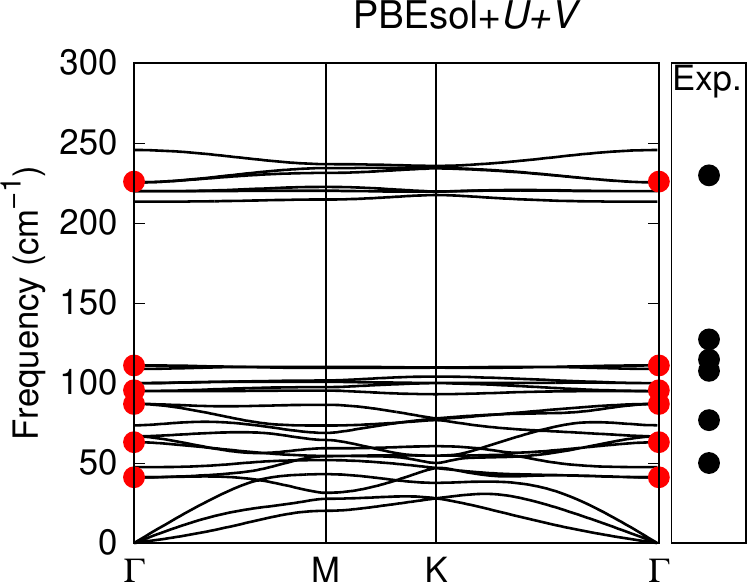}
             \vspace{-4ex}
             \label{cr-ph-3}
    \end{subfigure}
    \newline
    \begin{subfigure}[t]{.33\textwidth}
              \centering
             \caption{}
             \includegraphics[scale=0.75]{ 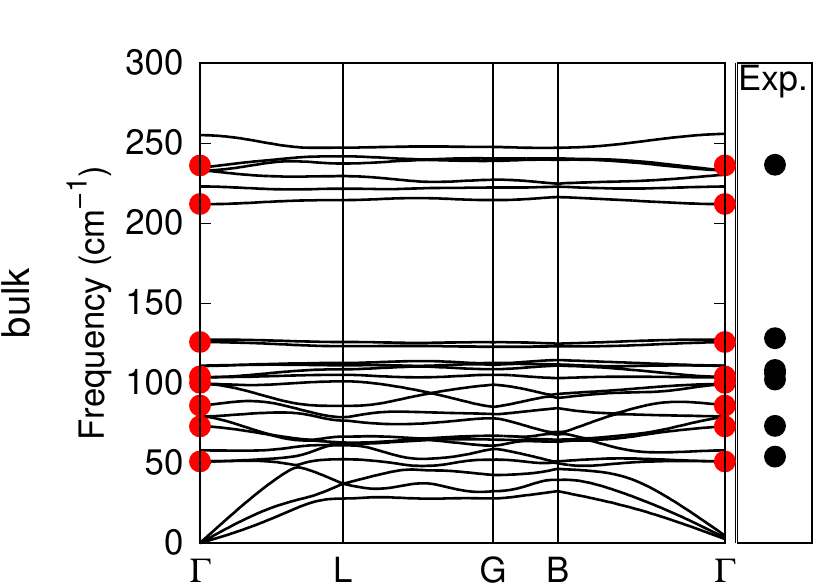}
             \vspace{-4ex}
             \label{cr-ph-4}
    \end{subfigure}
    \hfill
    \begin{subfigure}[t]{.3\textwidth}
             \centering
             \caption{}
             \includegraphics[scale=0.75]{ 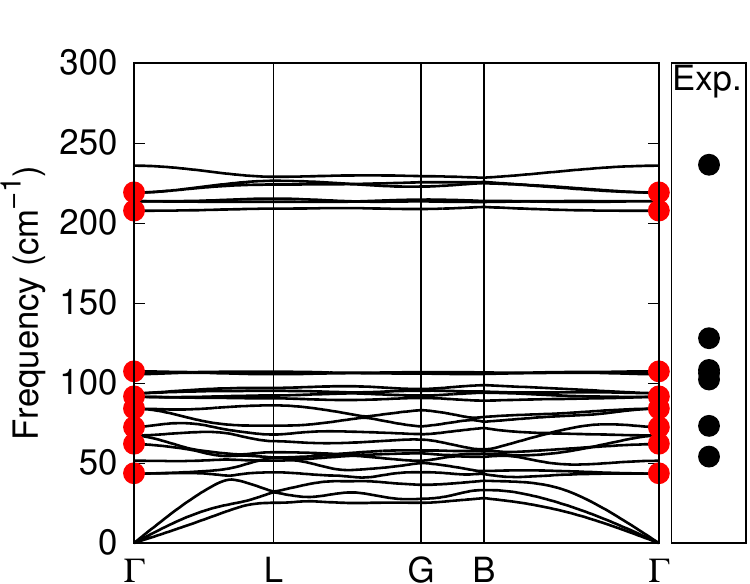}
             \vspace{-4ex}
             \label{cr-ph-5}
    \end{subfigure}
    \hfill
    \begin{subfigure}[t]{.3\textwidth}
             \centering
             \caption{}
             \includegraphics[scale=0.75]{ 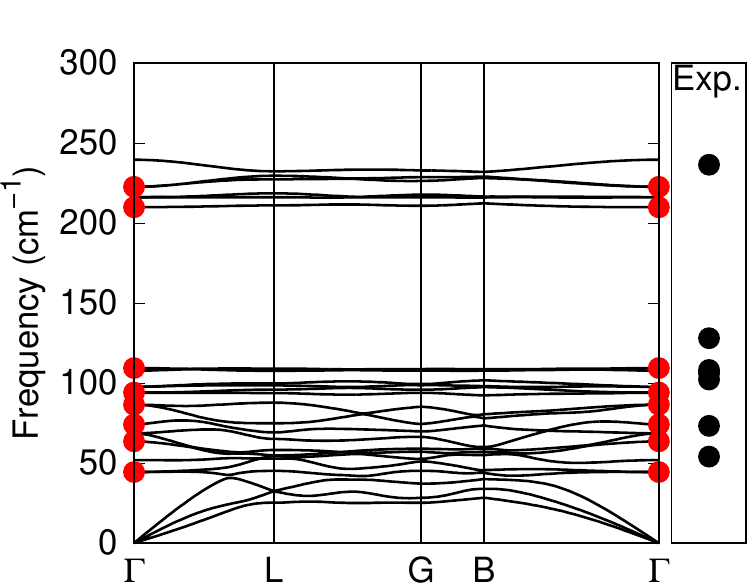}
             \vspace{-4ex}
             \label{cr-ph-6}
    \end{subfigure}
    \captionsetup{justification=Justified}
    \caption{phonon bands  of monolayer (first row) and bulk (second row) of CrI$_3$ from (a) and (d) PBEsol, (b) and (e) PBEsol+\emph{U} and (c) and (f) PBEsol+\emph{U}+\emph{V}. The frequencies at $\Gamma$ point for the Raman active modes are shown by red dots compared to the Raman experiments data  by black dots.}
    \label{cr-ph-band}
\end{figure*} 

As it can be seen in \cref{cr-band}, adding the Hubbard corrections shifts up the Cr($3d$) spin minority conduction bands by 3~eV. This makes the spacing between the spin-majority and the spin-minority conduction bands much larger than that reported from analysis of scanning tunneling spectroscopy (STS) experiments for few-layer CrI$_3$  (0.8~eV)~\cite{qiu_visualizing_2021}. As a consequence, one might infer that for CrI$_3$ it is better to neglect the effects of Hubbard interactions since PBEsol already shows good agreement with experiments. However, ARPES experiments suggest that the top of the valence bands is dominated by I($5p$) states~\cite{kundu_valence_2020}, a feature that  PBEsol fails to reproduce but Hubbard corrections improve by pushing the Cr($3d$) states down to lower energies. This begs the question: How can one correct the valence band edge character while not adversely affecting the spin-minority conduction bands?

As already reported elsewhere~\cite{qiu_visualizing_2021,jiang_spin_2018}, the relative position of the spin-minority conduction bands strongly depends on the value of the \emph{U} parameter, with a larger shift of the spin-minority bands with increasing \emph{U}. The effect is particularly dramatic in \cref{cr-band}, given the large value of $U=6.54$~eV, although we mention that in Ref.~\cite{liu_exfoliating_2016} the authors calculated \emph{U} from the linear-response method~\cite{kulik_density_2006} to be only 2.65~eV for CrI$_3$ monolayer. Meanwhile, in Ref.~\cite{qiu_visualizing_2021} the authors reproduce the experimental splitting between the spin-majority and spin-minority conduction states by using a small empirical value of  $U=0.5$~eV. Finally, Ref.~\cite{sarkar_electronic_2020-1} found that DFT+\emph{U} can correct the large splitting between the spin-majority/spin-minority conduction bands, if the around-mean-field (AMF) double-counting formulation is used. (In this work, we opt to focus on the FLL double-counting scheme, because the resulting DFT+$U$ functional is a tailored correction to address piece-wise linearity.)
 
We address the incorrect positioning of the spin-minority conduction bands in an alternative and non-empirical way, by investigating the effect of a different Hubbard \emph{U} for the two spin channels. We calculate the spin-resolved \emph{U} using the approach of Ref.~\cite{linscott_role_2018}, which is more appropriate when the two spin channels are not strongly coupled and we want to linearize the total energy with respect to the inter-spin-channel density response. Consequently, the off-diagonal elements of the response functions are not considered. More details about the calculation of a spin-resolved \emph{U} are provided in Supplemental Material. The calculated self-consistent spin-resolved $U^\sigma$ for spin-up and spin-down channels for the monolayer are $U^{\uparrow}=1.72$~eV and $U^{\downarrow}=0.31$~eV. We notice $U^{\downarrow}$ is an order of magnitude smaller than the conventional \emph{U} given in \cref{t3}. The band structure of CrI$_3$ monolayer from PBEsol+$U^{\uparrow}$+$U^{\downarrow}$ compared with PBEsol+\emph{U} is illustrated in \cref{spin-res-band}. For these spin-resolved \emph{U} calculations, the Cr($3d$) spin-minority conducting states are in a correct position and the top of the valence bands are mainly dominated by I($5p$) states, consistent with ARPES and STS experiments. We will see later that a spin-resolved \emph{U} can also improve the vibrational frequencies (\cref{t4}), thus making the approach particularly promising. We note in passing that we also tested the inclusion of Hund's exchange parameter $J$~\cite{himmetoglu_first-principles_2011}. The results are summarized in the Supplemental Material, showing that $J$ only marginally improves the band structure and can introduce additional artifacts.

\begin{table*}[ht!]
    \caption{\label{t4}  The Raman active modes for CrI$_3$ monolayer and bulk systems computed with PBEsol, PBEsol+$U$, PBEsol+$U$+$V$ and PBEsol+$U_\uparrow$+$U_\downarrow$ and compared with experiments.}
    \centering
    \renewcommand{\arraystretch}{1.5}
    \begin{tabularx}{\textwidth}{ X X D..{6.4} *{4}{D..{5.3}} *{4} {D..{6.3}}{D..{5.3}}}
    \hline
    \hline
    &
    &\multicolumn{1}{c}{E$_g$} 
    &\multicolumn{1}{c}{A$_g$} 
    &\multicolumn{1}{c}{E$_g$} 
    &\multicolumn{1}{c}{E$_g$} 
    &\multicolumn{1}{c}{A$_g$} 
    &\multicolumn{1}{c}{A$_g$} 
    &\multicolumn{1}{c}{A$_g$} 
    &\multicolumn{1}{c}{E$_g$}  \\
    \hline
    \multirow{4}{*}{bulk}
    & Expt.                      & 54.1 & 73.3 & 102.3 & 106.2 & 108.3 & 128.1 & \multicolumn{1}{c}{---} & 236.6 \Tstrut \\[0.1cm]
    & PBEsol                     & 51   & 73   & 100   & 104   & 86    & 126   & 212   & 236 \Tstrut \\[0.1cm]
    & PBEsol+\emph{U}            & 44   & 62   & 84    & 92    & 72.5  & 107   & 208   & 219 \\[0.1cm]
    & PBEsol+\emph{U}+\emph{V}   & 44   & 64   & 86.5  & 94    & 74    & 109   & 210   & 223 \\[0.1cm]
    \hline
    \hline
    &
    &
    &\multicolumn{1}{c}{E$_g$ } 
    &\multicolumn{1}{c}{A$_{1g}$} 
    &\multicolumn{1}{c}{E$_g$} 
    &\multicolumn{1}{c}{E$_g$} 
    &\multicolumn{1}{c}{A$_{1g}$} 
    &\multicolumn{1}{c}{A$_g$} 
    &  \\
    \hline
    \multirow{5}{*}{monolayer}
    & Expt.~\cite{huang_tuning_2020}       &  & 50  & 76.9 & 107.7 & 114.8 & 127.4 & 230   &  \Tstrut  \\[0.1cm]
    & PBEsol                               &  & 48  & 71.5 & 100   & 105   & 125.5 & 238   & \\[0.1cm]
    & PBEsol+\emph{U}                      &  & 40  & 62   & 85    & 93    & 109   & 223   &  \\[0.1cm]
    & PBEsol+\emph{U}+\emph{V}             &  & 41  & 63   & 87    & 95.5  & 111   & 226   & \\[0.1cm]
    & PBEsol+$U^\uparrow$+$U^\downarrow$   &  & 48  & 71.4 & 99    & 105   & 127   & 238 \\[0.1cm]
    \hline
    \hline
    \end{tabularx}
\end{table*}

Next, we study the vibrational properties of CrI$_3$. The calculated dielectric tensor and BECs are reported in~\cref{t4} for the bulk systems. In the case of PBEsol+$U$, the dielectric tensor is increased considerably, indicating the effect of the Hubbard correction in localizing electrons. The phonon dispersion and the corresponding frequencies at the $\Gamma$ point are shown in \cref{cr-ph-band} and summarized in \cref{t4} for the Raman active modes. We note that in a recent study~\cite{bonini_frequency_2023} the authors show that the effects arising from broken time-reversal symmetry in the interatomic force constants (IFC) split the two-fold degenerate $E_g$ (and $E_u$) modes at $\Gamma$ into chiral modes, albeit with a very small splitting. The frequencies in \cref{t4} are obtained with conventional IFC, thus preserving the degeneracy of $E_g$ modes. It is evident from \cref{t4} that the PBEsol results agree well with experiments and Hubbard corrections seem to worsen the comparison. Hubbard \emph{V} corrections  improve upon PBEsol+\emph{U} but still do not perform as well as PBEsol. This poor agreement is  attributed to the strength of the Hubbard \emph{U} from linear response within DFPT, which can affect the structural optimization, leading in particular to an overestimation of the lattice parameter (see \cref{t2}) and thus a softening of the phonon frequencies. The results improve when employing a spin-resolved~\emph{U}, with an accuracy approaching that of PBEsol. Given that PBEsol poorly describes the valence electronic bands, we conclude that PBEsol+$U^{\uparrow}$+$U^{\downarrow}$ gives the best agreement with experiments across both electronic and vibrational properties.

\begin{table}[h]
    \caption{\label{t9} Dielectric tensor and BECs (for symmetry inequivalent atoms) of CrI$_3$ bulk from PBEsol and PBEsol+\emph{U} calculations in the Cartesian framework. The results from PBEsol+$U$ are used for the case of PBEsol+$U$+$V$ as well.}
    \centering
    \renewcommand{\arraystretch}{1.5}
    \begin{tabularx}{\columnwidth}{c c c }
    \hline
    \hline
    &  PBEsol  &  PBEsol+\emph{U} \\
    \hline
    $\epsilon^{\infty}$ & 
    $\begin{pmatrix}  
        8.46   & 0.0   & 0.0 \\
        0.0    & 8.46  & 0.0 \\
        0.0    & 0.0   & 6.33 
    \end{pmatrix}$ 
    & 
    $\begin{pmatrix}  
        11.42  & 0.0    & 0.0 \\
        0.0    & 11.42  & 0.0 \\
        0.0    & 0.0    & 9.89  
    \end{pmatrix}$\\ 
    \hline 
    Z$_{Cr}$ &  
    $\begin{pmatrix} 
         2.43  & 0.03   & 0.0 \\
        -0.03  & 2.43   & 0.0 \\
         0.0   & 0.0    & 0.90 
    \end{pmatrix}$ 
    & 
    $\begin{pmatrix} 
         1.60  & 0.04   & 0 \\
        -0.04  & 1.60   & 0  \\
         0.0   & 0.0    & 0.49  
    \end{pmatrix}$ \\ 
    \hline
    Z$_{I}$ &  
    $\begin{pmatrix}  
        -0.32  & -0.03  & -0.41 \\
        -0.01  & -1.31  & -0.05 \\
        -0.27  & -0.10  & -0.30 
    \end{pmatrix}$ 
    & 
    $\begin{pmatrix} 
        -0.26  & -0.06  & -0.29 \\
        -0.03  & -0.81  & -0.17 \\
        -0.10  & -0.23  & -0.16 
    \end{pmatrix}$ \\
    \hline
    \hline
    \end{tabularx}
\end{table}

%-----------------------------------------------------------------------------

\section{Conclusions}
\label{sec4}

In conclusion, we study and benchmark the structural, vibrational, and electronic properties of FePS$_3$ and CrI$_3$ monolayers, which are representative of the growing class of 2D magnets, as well as their corresponding bulk counterparts. Our calculations make use of PBEsol and Hubbard corrected PBEsol (PBEsol+\emph{U}) and its extension (PBEsol+\emph{U}+\emph{V}). The on-site (\emph{U}) and inter-site (\emph{V}) Hubbard interactions are calculated within DFPT. For the case of FePS$_3$, Hubbard corrections play a crucial role in describing the insulating ground state of the system with the correct experimental symmetry, while at the PBEsol level the system is metallic and develops phonon instabilities that drive it towards a lower symmetry state. The case of CrI$_3$ is more complex, as PBEsol calculations already provide good structural and vibrational properties that seem to be worsen by Hubbard corrections. Still, PBEsol+\emph{U}(+\emph{V}) is needed to describe correctly the orbital content of the top valence bands, although it gives rise at the same time to a spurious shift in the spin-minority conduction bands. Using a spin-resolved~\emph{U} recovers a correct description of both the valence and conduction bands, together with excellent structural and vibrational properties, providing the best overall agreement with experiments.

%-----------------------------------------------------------------------------------
\acknowledgments
\section{acknowledgments}
We thank Changpeng Lin, Michele Kotiuga, and Lorenzo Bastonero for fruitful discussions. We acknowledge support from the Swiss National Science Foundation (SNSF), through grant 200021-179138, and its National Centre of Competence in Research (NCCR) MARVEL (grant number 205602). M.G.\ acknowledges financial support from the Italian Ministry for University and Research through the Levi-Montalcini program and through the PNRR project ECS\_00000033\_ECOSISTER. Computer time was provided by the Swiss National Supercomputing Centre (CSCS) under project No.~s1073.

\clearpage 
\bibliographystyle{apsrev4-1}

\end{document}